\documentclass[12pt]{article}
\usepackage[dvips]{graphicx}
\usepackage{epsfig}
\usepackage{amsmath,amssymb,amsbsy}
\usepackage{amsfonts}
\usepackage{color}
\usepackage{soul}

\begin{document}

\begin{flushright}
LU TP 15-33\\
March 2016
\vskip1cm
\end{flushright}

\thispagestyle{empty}
\begin{center}
{\Large\bf{
The Drell-Yan process as a testing ground for parton distributions up to LHC}}
\vskip0.4cm
{\bf Eduardo Basso}
\vskip 0.2cm
Department of Astronomy and Theoretical Physics, \\
Lund University, SE 223-62 Lund, Sweden\\
\vskip 0.4cm
{\bf Claude Bourrely}
\vskip 0.2cm
Aix Marseille Universit\'e, Universit\'e de Toulon, CNRS, CPT, UMR 7332\\ 
13288 Marseille, Cedex 09, France\\
\vskip 0.4cm
{\bf Roman Pasechnik}
\vskip 0.2cm
Department of Astronomy and Theoretical Physics, \\
Lund University, SE 223-62 Lund, Sweden\\
\vskip 0.4cm
{\bf Jacques Soffer}
\vskip 0.2cm
Physics Department, Temple University,\\
1925 N, 12th Street, Philadelphia, PA 19122-1801, USA
\vskip 0.5cm
{\bf Abstract}\end{center}

The Drell-Yan massive dilepton production in hadron-hadron collisions provides
a unique tool, complementary to Deep Inelastic Scattering, for improving our
understanding of hadronic substructure and in particular for testing parton
distributions. We will consider measurements of the differential and
double-differential Drell-Yan cross sections from FNAL Tevatron up to CERN LHC
energies and they will be compared to the predictions of perturbative QCD
calculations using most recent sets (CT14 and MMHT14) of parton distribution 
functions, as well as those provided by the statistical approach.

\vskip 0.3cm

\noindent {\it Key words}:  Drell-Yan process, parton distribution functions \\

\noindent PACS numbers: 12.40.Ee, 13.38.Bx, 14.70.Hp

\newpage 

\section{Introduction}
\setcounter{page}{1}
The Deep Inelastic Scattering (DIS) of leptons and nucleons is indeed our main source 
of information to study the internal nucleon structure in terms of parton distribution 
functions (PDF). However at hadron colliders, one has access to the Drell-Yan (DY) 
process \cite{dy} and the measurement of massive dilepton production, via $Z/\gamma^*$ 
exchange in the $s$ channel $Z/\gamma^* \to l\bar l$, is also an excellent, unique 
and clean observable for testing QCD and, in particular, the nucleon structure encoded in PDFs \cite{Peng}. 
Needless to mention, the DY process is one of the standard candles for New Physics searches at the energy and luminosity frontiers 
which thus requires a significant reduction in theoretical uncertainties. This is the reason why several DY experiments 
were already performed in almost all high-energy hadronic facilities and they are now being intensively studied 
in all on-going major experiments at the Large Hadron Collider (LHC) at CERN. This will allow us 
to test various theoretical models in a vast kinematical domain and to analyse the corresponding PDF 
uncertainties with this simple process whose mechanism is dominated by parton-antiparton annihilation 
in $pp$ collisions.

The rapidity $y$ and the invariant mass $M_{l\bar l}$ of the dilepton system produced 
in proton-proton collisions are related, at leading order (LO), to the momentum fraction $x_+(x_-)$ 
carried by the parton in the forward-going (backward-going) proton, according to the formula 
$x_{\pm} =(M_{l\bar l}/\sqrt{s}){e}^{\pm y}$. Therefore, the rapidity and mass distributions are sensitive 
to the PDFs of the interacting partons. Besides, the transverse momentum $p_T$ distributions of the dilepton 
provides an additional information about the dynamics of proton collisions at high energy 
particularly sensitive to the higher-order QCD corrections. In this work, several most
recent PDFs parameterizations at the next-to-leading order (NLO) such as MMHT14 \cite{mmht14} and 
CT14~\cite{ct14} models, as well as the NLO statistical PDFs previously developed in Ref.~\cite{bs15}, will be employed
for description of the existing DY data for the differential $y$, $M_{l\bar l}$ and $p_T$ distributions, 
both at low and high energies. For this purpose, we have selected the data from a limited number 
of experiments at FNAL Tevatron (E866, D0) and CERN LHC (CMS, ATLAS) keeping only those with the
highest integrated luminosity, corresponding to the highest precision.

The paper is organized as follows. In Section 2, we review the main features of three sets of PDFs we 
have used for our calculations. In Section 3, we consider the invariant mass distribution at LHC (7/8 TeV) 
energies vs available data over a very broad dilepton mass range up to 2 TeV, and also in 
a much smaller mass range from a fixed-target FNAL experiment. In Section 4, the differential cross section 
as a function of the $Z/\gamma^*$ rapidity is analyzed. In Section 5, we study 
the $Z/\gamma^*$ transverse momentum spectra for two different centre-of-mass energies. 
We give our final remarks and conclusions in Section 6.


\section{PDFs selection}

We will now summarize the essential properties of three sets of PDFs which will be tested in our 
analysis of the DY process.

First, let us recall the main features of the statistical approach \cite{bs15} for building up the PDFs as opposed
to the standard polynomial type parameterizations based on Regge theory at 
low $x$ and on counting rules at large $x$. The fermion distributions are given by the sum of 
two terms, a quasi Fermi-Dirac function and a helicity independent diffractive
contribution
\begin{equation}
xq^h(x,Q^2_0)=
\frac{A_{q}X^h_{0q}x^{b_q}}{\exp [(x-X^h_{0q})/\bar{x}]+1}+
\frac{\tilde{A}_{q}x^{\tilde{b}_{q}}}{\exp(x/\bar{x})+1}~,
\label{eq1}
\end{equation}
\begin{equation}
x\bar{q}^h(x,Q^2_0)=
\frac{{\bar A_{q}}(X^{-h}_{0q})^{-1}x^{b_{\bar q}}}{\exp [(x+X^{-h}_{0q})/\bar{x}]+1}+
\frac{\tilde{A}_{q}x^{\tilde{b}_{q}}}{\exp(x/\bar{x})+1}~,
\label{eq2}
\end{equation}
defined at the input energy scale $Q_0^2=1\,\mbox{GeV}^2$. We note that the diffractive 
term is absent in the quark helicity distribution $\Delta q$ and in the quark valence contribution 
$q - \bar q$.

In Eqs.~(\ref{eq1}) and (\ref{eq2}) the multiplicative factors $X^{h}_{0q}$ and $(X^{-h}_{0q})^{-1}$ 
in the numerators of the non-diffractive parts of the $q$'s and $\bar{q}$'s distributions, imply a modification 
of the quantum statistical form, which was proposed in order to agree with experimental data \footnote{These factors 
were fully justified in the extension of the PDFs to include their transverse momentum dependence (TMD) \cite{bbs13}.}. 
The parameter $\bar{x}$ plays the role of a {\it universal temperature} and $X^{\pm}_{0q}$ are the two {\it thermodynamical potentials} 
of the quark $q$, with helicity $h=\pm$. They represent the fundamental characteristics of the model. 
Notice the change of sign of the potentials and helicity for the antiquarks\footnote{At variance with statistical 
mechanics where the distributions are expressed in terms of the energy, here one uses
$x$ which is clearly the natural variable entering in all the sum rules of the parton model.}.

\begin{figure}[hbt]  
\begin{center}
\includegraphics[width=6.0cm]{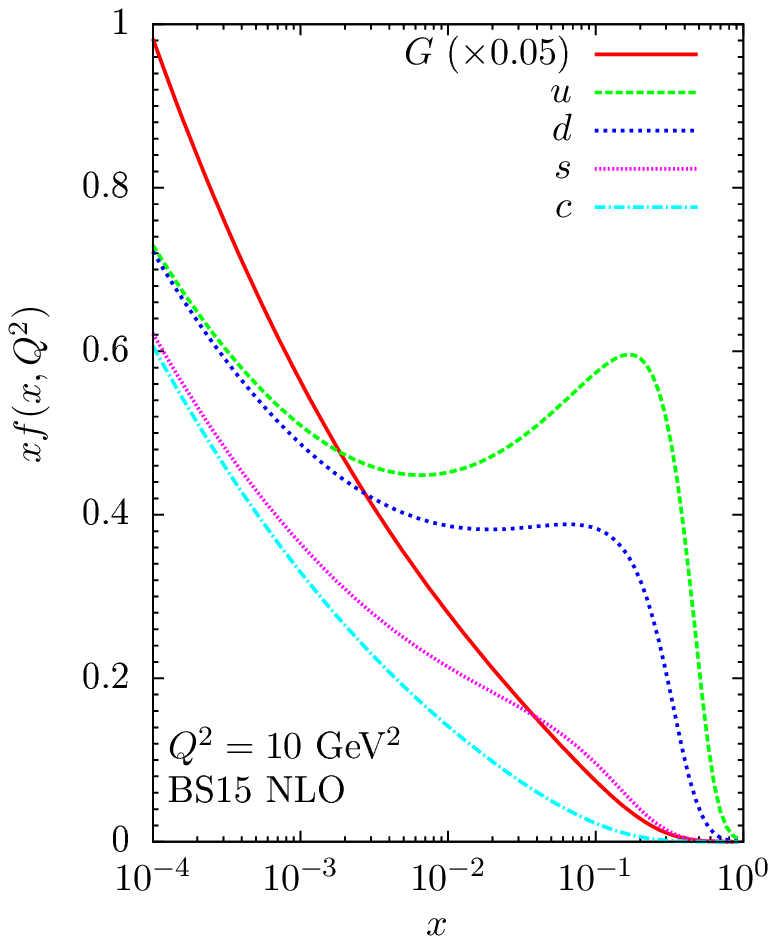}
\includegraphics[width=6.0cm]{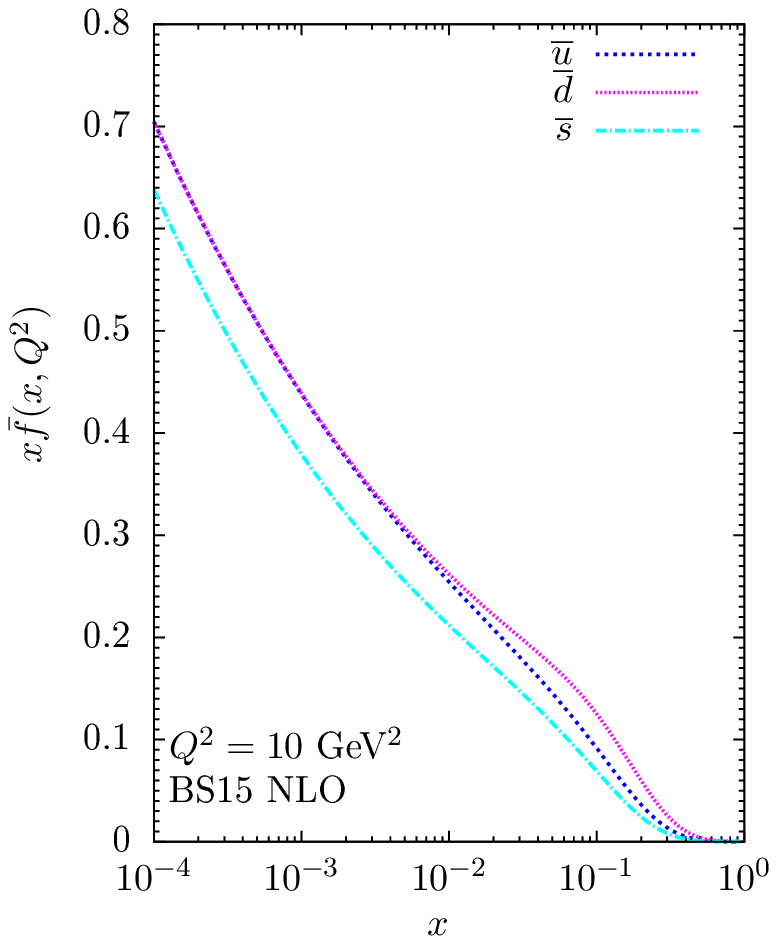}
\caption[*]{\baselineskip 1pt
The BS15 PDFs for quarks, gluon ($\it{left}$) and antiquarks ($\it{right}$) from 
Ref.~\cite{bs15} at $Q^2 = 10\,\mbox{GeV}^2$.}
\label{bs15}
\end{center}
\end{figure}

Although the statistical approach to the starting PDFs allows the simultaneous description of unpolarized 
cross sections and helicity asymmetries, a unique situation in the literature, in this work we will restrain ourselves 
to spin-independent DY observables. For a given flavor $q$, the corresponding quark and antiquark distributions 
involve {\it eight} free parameters: $X^{\pm}_{0q}$, $A_q$, $\bar {A}_q$, $\tilde {A}_q$, $b_q$, $\bar {b}_q$ and $\tilde {b}_q$. 
It reduces to $\it seven$ since one of them is fixed by the valence sum rule, $\int (q(x) - \bar {q}(x))dx = N_q$, 
where $N_q = 2, 1, 0 ~~\mbox{for}~~ u, d, s$, respectively. In the light quark sector $q=\{u,d\}$, the total number 
of free parameters is reduced to $\it eight$ by applying additional constraints 
as was done in Ref. \cite{bbs1} (for a more detailed review, see e.g. Ref.~\cite{bbs-rev})
\[ A_u=A_d\,,\quad  \bar {A}_u = \bar {A}_d\,, \quad \tilde {A}_u = \tilde {A}_d\,, \quad b_u = b_d\,, \quad 
\bar {b}_u = \bar {b}_d\,, \quad \tilde {b}_u = \tilde {b}_d \,. \] 
For the gluon PDF at the input energy scale, we consider the black-body 
inspired expression
\begin{equation}
xG(x,Q^2_0) = \frac{A_Gx^{b_G}}{\exp(x/\bar{x})-1}~,
\label{eq3}
\end{equation}
a quasi Bose-Einstein function, with $b_G$ being the only free parameter, since $A_G$ is determined 
by the momentum sum rule. To summarize, this determination of all PDF sets\footnote{In Ref.~\cite{bs15} 
we have considered the helicity gluon distribution which is irrelevant in the present work} involves a total of {\it seventeen} 
free parameters. Namely, in addition to the temperature $\bar x$ and the exponent $b_G$ of the gluon distribution, 
we have {\it eight} free parameters for the light quarks $(u,d)$, {\it seven} free parameters for the strange quarks as was
outlined above. These parameters were determined in Ref.~\cite{bs15}, from a NLO QCD fit of 
a large set of accurate DIS data ${\it only}$. The resulting PDFs, denoted from now on as BS15, are illustrated for the 
energy scale $Q^2 = 10\,\mbox{GeV}^2$ in Fig.~\ref{bs15}.


Next we consider an improved version of the so-called MSTW framework which was proposed seven years ago in Ref.~\cite{mstw08}. 
The main shortcoming of the MSTW PDFs was an incorrect description of the lepton charge asymmetry from $W^{\pm}$ decays as a function 
of the lepton rapidity. The new input distributions result from major changes in the theoretical procedure with respect to original Ref.~\cite{mstw08} 
since they involve Chebyshev polynomials. In this new version proposed in Ref.~\cite{mmht14} the majority of the starting PDFs 
have the following form
\begin{equation}
xf(x,Q_0^2) = A(1 - x)^{\eta}x^{\delta}[ 1 + \sum_{i=1}^{n}a_i T^{Ch}_i (y(x))]~,
\end{equation}
where $Q_0^2 = 1 \mbox{GeV}^2$ is the input energy scale, and $T^{Ch}_i (y)$ are Chebyshev polynomials in $y$, with
$y = 1 - 2\sqrt{x}$, where one takes $n = 4$. For each PDF, namely, $f = u_V,\, d_V,\, S,\, s_+$ one has to determine the parameters 
$A,\, \eta,\, \delta,\, a_i$. Here, $u_V$ and $d_V$ denote the light-quark valence distributions and $S \equiv 2(\bar u + \bar d) + s_+$ is the light-quark 
sea distribution. For $s_+ \equiv s +\bar s$ one sets $\delta_+ = \delta_S$.
\begin{figure}[hbt]  
\begin{center}
\includegraphics[width=6.0cm]{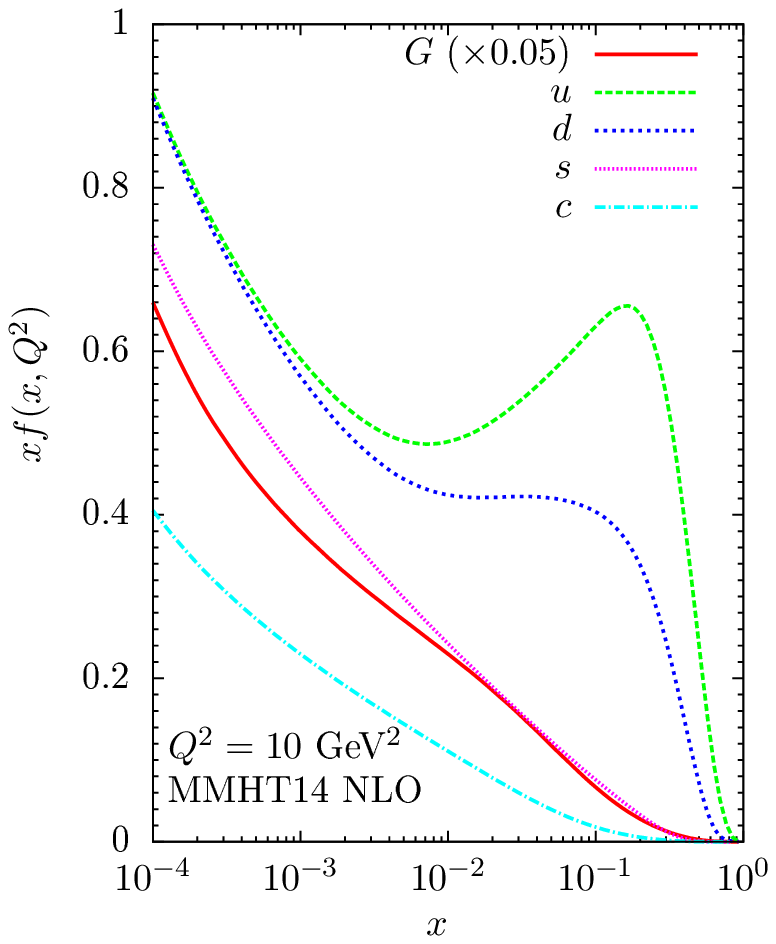}
\includegraphics[width=6.0cm]{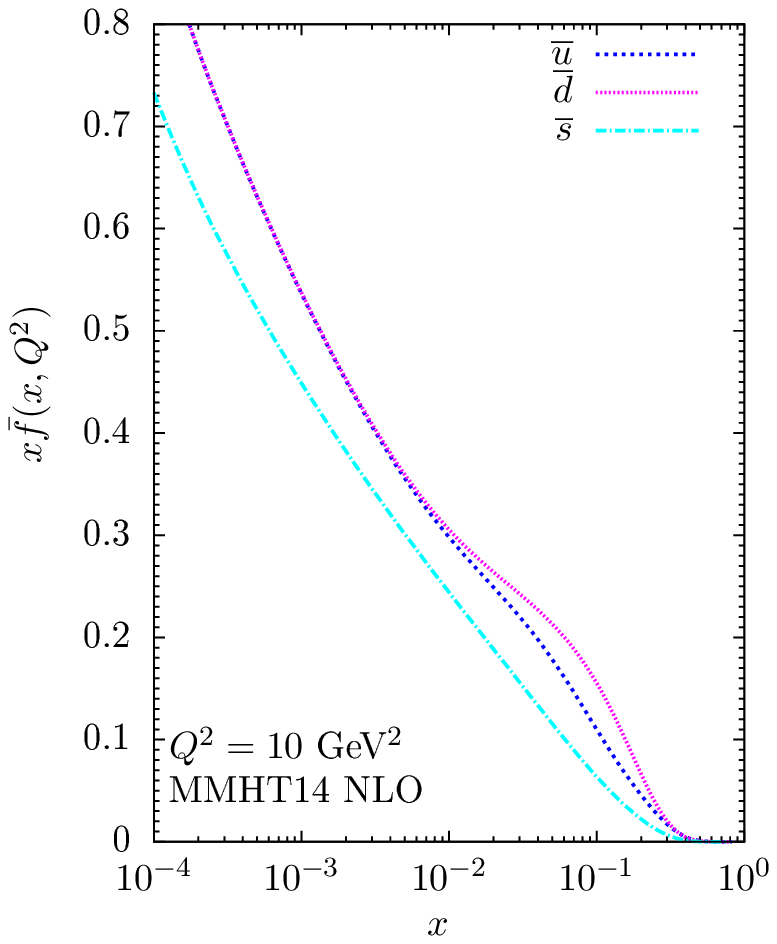}
\caption[*]{\baselineskip 1pt
The MMHT14 PDFs for quarks, gluon ($\it{left}$) and antiquarks ($\it{right}$) from 
Ref.~\cite{mmht14} at $Q^2 = 10\, \mbox{GeV}^2$.}
\label{mmht14}
\end{center}
\end{figure}

We still have to specify the parameterisations of the gluon and of the differences $\Delta \equiv \bar d - \bar u$ and $s - \bar s$. 
For $\Delta$ one sets $\eta_{\Delta} = \eta_S + 2$ and one uses the following expression
\begin{equation}
x\Delta(x,Q_0^2) = A_{\Delta}(1 - x)^{\eta_{\Delta}}x^{\delta_{\Delta}}( 1 + \gamma_{\Delta} x + \epsilon_{\Delta} x^2)~.
\end{equation}
For the poorly determined strange quark difference one takes
\begin{equation}
s_- \equiv x(s - \bar s) = A_- ( 1 - x )^{\eta_-} x^{\delta_-} (1 - x/x_0)~.
\end{equation}
Finally, for the gluon distribution, as proposed long time ago \cite{mrst02}, one needs a second term for the small 
$x$ behavior as shown below
\begin{equation}
xG(x,Q_0^2) = A_G(1 - x)^{\eta_G}x^{\delta_G}[ 1 + \sum_{i=1}^{2}a_{G,i} T^{Ch}_i (y(x))] + A_{G^{'}}(1 - x)^{\eta_{G^{'}}} x^{\delta_{G^{'}}}~,
\end{equation}
and we notice that it involves {\it seven} free parameters since $A_G$ is constrained by the momentum sum rule. 
This is a major difference with respect to the statistical approach. In total, there are {\it thirty seven} free parameters, 
a large number, while one should remember that there are three constraints from the valence sum rules as 
already mentioned above. In addition, there is also the strong coupling defined at the mass scale of the $Z$ boson, i.e. $\alpha_s(M_Z^2)$, 
which is allowed to be free when determining the best fit. The authors claim that the advantage of using a parameterisation based on 
Chebyshev polynomials is the stability and a good convergence of the values found for the coeffients $a_i$. In what follows, we refer
to these PDFs as to the MMHT14 NLO model.
\begin{figure}[hbt]  
\begin{center}
\includegraphics[width=6.5cm]{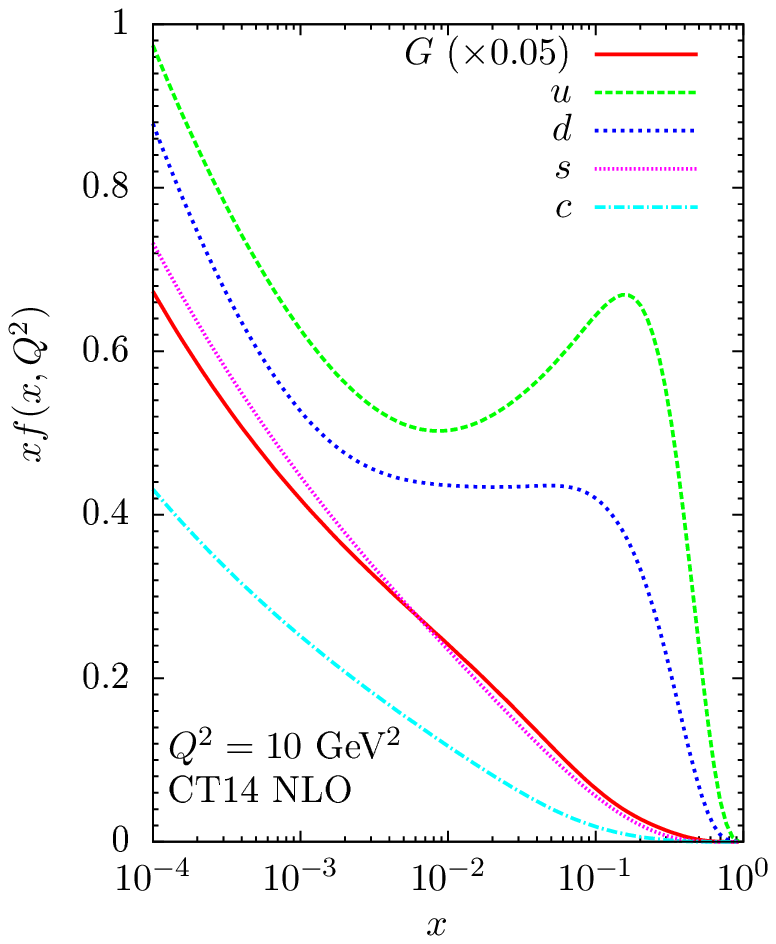}
\includegraphics[width=6.5cm]{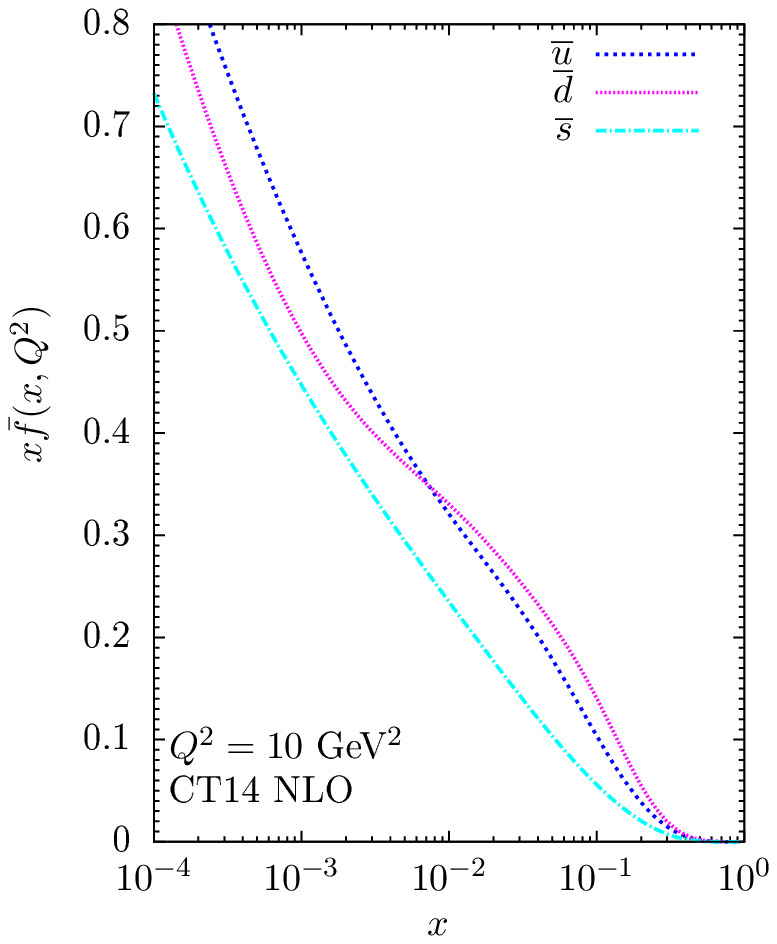}
\caption[*]{\baselineskip 1pt
The CT14 PDFs for quarks, gluon ($\it{left}$) and antiquarks ($\it{right}$) from Ref.~\cite{ct14} 
at $Q^2 = 10\,\mbox{GeV}^2$.}
\label{ct14}
\end{center}
\end{figure}

The parameters were determined from a global analysis of a variety of new data sets, from the LHC including some DY results, updated 
Tevatron data and HERA combined H1 and ZEUS data on the total and charm structure functions. For DIS data on deuterium targets deuteron 
corrections were taken into account, as well as nuclear corrections for neutrino data. The NLO and a next-to-next-to-leading order (NNLO) QCD fits 
of the data were performed and we display  the resulting PDFs in Fig.~\ref{mmht14} at the same energy scale $Q^2 = 10\,\mbox{GeV}^2$ as in Fig.~\ref{bs15}. 
The fit quality and the determination of the PDF uncertainties were carefully studied in Ref.~\cite{mmht14}. It is interesting to observe that 
in most cases the new MMHT PDFs are within one standard deviation from those of the MSTW framework \cite{mstw08}, a remarkable stability 
of the procedure.

We now turn to the third PDF set we will use in our calculations of the DY cross sections. It results from different versions \cite{cteq} 
up to NNLO, from the CTEQ-TEA global analysis of QCD, including the use of data from HERA, Tevatron and LHC. New theoretical developments 
are associated to the achievement of an increasing data precision in DIS, vector boson production and single-jet production. It is interesting 
to note that DY data from colliders were not considered so far. Standard parametrizations for each flavor are of the form
\begin{equation}
x f_a(x,Q_0^2)  =  x^{a_1} (1-x)^{a_2} P_a(x) \,,
\label{xfa}
\end{equation}
where the first two factors are suggested by the Regge theory and by counting rules, and the remaining factor $P_a(x)$ 
is assumed to be slowly varying. In the previous CTEQ analyses, $P_a(x)$ for each flavor was chosen as an exponential 
of a polynomial in $x$ or $\sqrt{x}$, for instance,
\begin{equation}
P_{q_v}(x) = \exp(a_0 + a_3\sqrt{x} + a_4 x + a_5 x^2) \,,
\label{qv}
\end{equation}
for valence quarks $q_v$.

In their most recent work refered to as CT14 \cite{ct14}, for the valence quarks 
they re-express the polynomial as a linear combination of \emph{Bernstein polynomials} in $y=\sqrt{x}$
\begin{equation}
P_{q_v} \, =  d_0  p_0(y)  +  d_1  p_1(y)  + d_2  p_2(y)  +  d_3  p_3(y)  +  d_4  p_4(y) \,,
\end{equation}
where $p_0(y)=(1-y)^4, ~p_1(y)=4 y (1 - y)^3, ~p_2(y)=6 y^2 (1 - y)^2, ~p_3(y) = 4  y^3  (1 - y) ~\mbox{and} ~p_4(y) = y^4$.
Then seven parameters for each flavor are reduced to just four by setting $d_1 = 1$, $d_3 = 1 + a_1/2$ and using 
the valence sum rule. So the valence quark $u_v$ and $d_v$ are determined in terms of a total of {\it eight} free parameters.

The CT14 model uses a similar parameterisation for the gluon but with a polynomial of a lower order since the data provide 
fewer constraints on the gluon distribution
\begin{equation}
P_{g}(y)  =  g_0  \left[e_0  q_0(y)  +  e_1 q_1(y)  + q_2(y) \right] \,,
\end{equation}
where $q_0(y) = (1 - y)^2, ~q_1(y) = 2 y  (1 - y),~\mbox{and}~ q_2(y) = y^2$.
However, instead of $y=\sqrt{x}$, it employs $y  =  1  -  (1-\sqrt{x})^2  =  2\sqrt{x} - x $. 
The momentum sum rule reduces the total number of parameters of the gluon distributions down to {\it five}.
The sea quark distributions $\bar{d}$ and $\bar{u}$ were parametrized using fourth-order polynomials in $y$ with the same
variable $y \, = \, 2\sqrt{x}\, - \, x$ that was used for the gluon. They assumed $\bar{u}(x)/\bar{d}(x) \to 1$ at $x \to 0$, which 
implies $a_1(\bar{u}) = a_1(\bar{d})$.

All in all, the CT14 model has {\it eight} free parameters associated with the valence quarks, {\it five} parameters associated 
with the gluon, and {\it thirteen} parameters associated with sea quarks, which in total amounts to {\it twenty six} fitting parameters.
The initial energy scale $Q_0$=1.295 GeV has been used to perform the NLO and NNLO QCD fits of a large set of data. We display the resulting 
PDFs in Fig.~\ref{ct14} at $Q^2 = 10\,\mbox{GeV}^2$ as a reference.

All the PDFs considered in this work were obtained from their correspondent LHAPDF6 \cite{lhapdf6} grids, which were used in our 
Monte-Carlo simulations described below\footnote{The LHADPF grids for the BS15 NLO model shall appear in online database soon.}.


\section{Dilepton invariant mass distribution}

\begin{figure}[h!]
\begin{center}
\includegraphics[width=10.0cm]{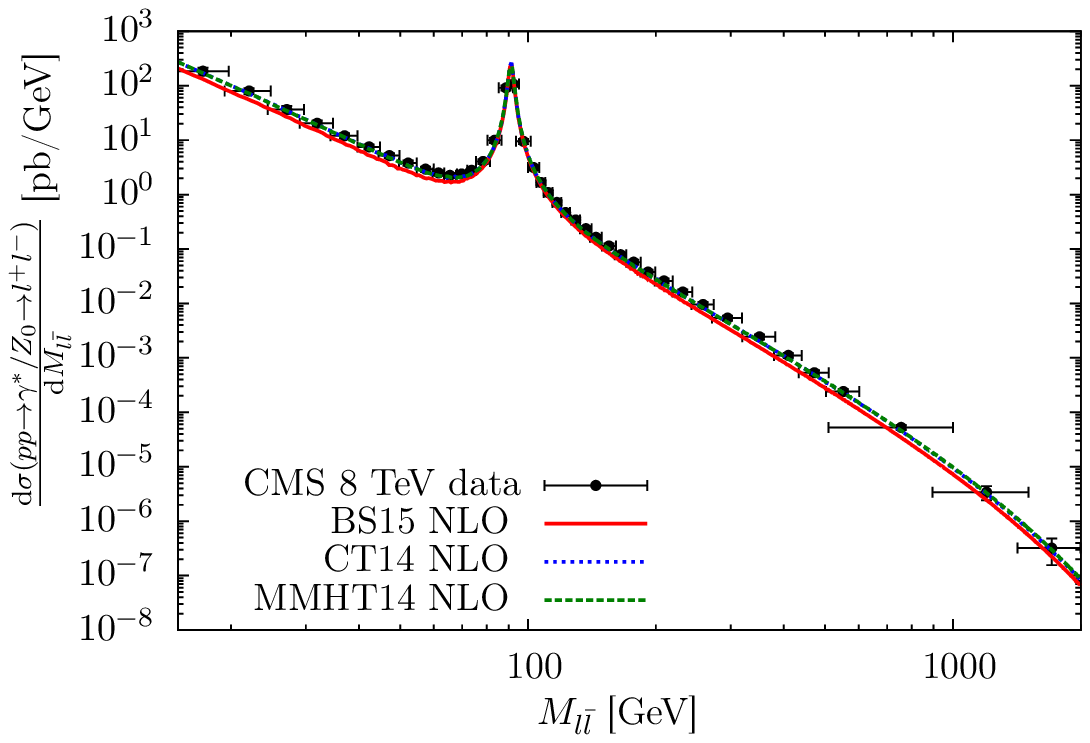}
\caption[*]{\baselineskip 1pt
The DY differential cross section measured in the combined di-muon and di-electron channels by CMS at $\sqrt{s}$ = 8 TeV 
over the invariant mass range from 15 GeV to 2 TeV \cite{cms15} vs QCD NLO predictions 
obtained by using PDF models from Refs.~\cite{mmht14,ct14,bs15}.}
\label{cms-8}
\end{center}
\end{figure}

Consider first, as displayed in Fig.~\ref{cms-8}, the measurement of DY cross section 
at the LHC energy $8$ TeV allowing to get dilepton mass up to 2 TeV.
The NLO cross section at LHC energies was calculated using the $q_T$ subtraction method implemented
in the DYNNLO code \cite{dynnlo}. This semi-analytical tool employs the dipole subtraction formalism of Catani and Seymor 
\cite{CSsubtract} realised by using the MCFM event generator \cite{mcfm} and stands as one of the 
up-to-date calculations of the Drell-Yan process up to NNLO. In the case of the invariant mass 
distributions the full phase space is integrated within each bin in the dilepton mass $M_{l\bar l}$. 
Note, in all the calculations here and below the factorisation 
and renormalisation scale are taken to be equal to the transverse mass of the dilepton pair, by convention. 
One notices that the theoretical results with distinct PDFs behave similarly for a broad range of the invariant 
mass with minor differences.
\begin{figure}[h!]
\begin{center}
\includegraphics[width=10.0cm]{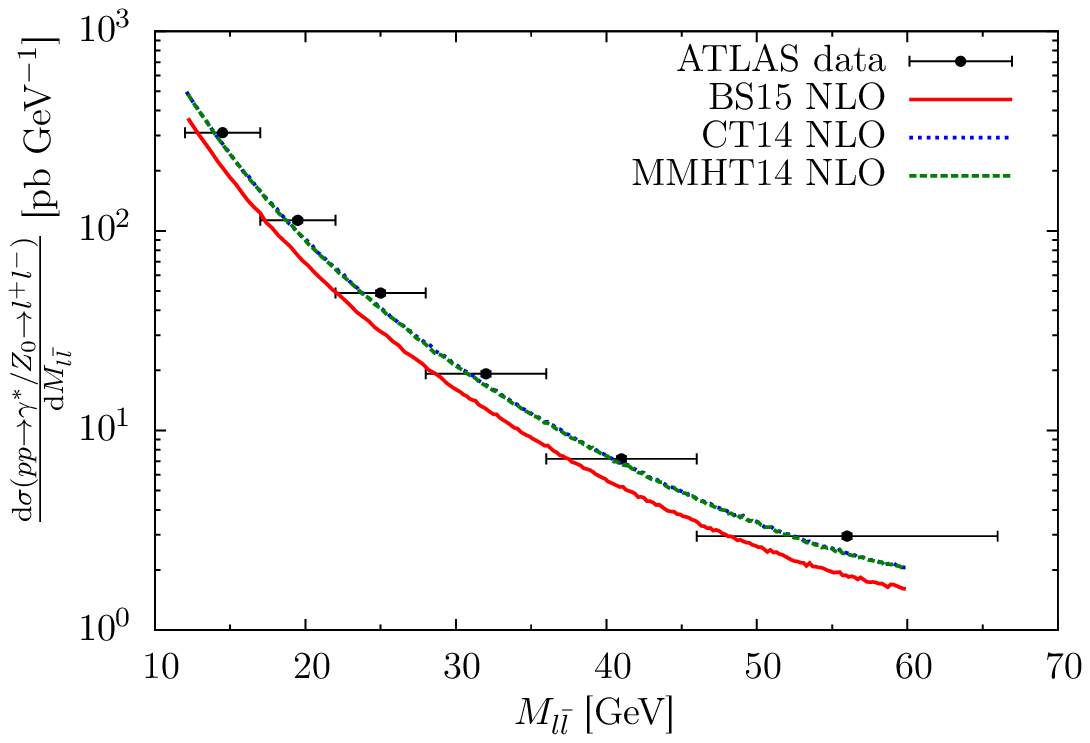}
\caption[*]{\baselineskip 1pt
The DY differential cross section measured in the combined di-muon and
di-electron channels by ATLAS at $\sqrt{s}$ = 7 TeV over the extended mass range 
12 -- 66 GeV \cite{atlas14_lowmass} vs QCD NLO predictions obtained by using 
PDF models from Refs.~\cite{mmht14,ct14,bs15}.}
\label{atlas-7-lomass}
\end{center}
\end{figure}

The DY low mass region was specifically measured by the ATLAS Collaboration at luminosities of 1.6 fb$^{-1}$ and 35 pb$^{-1}$ \cite{atlas14_lowmass}, 
for which a comparison with three distinct NLO PDFs based on the DYNNLO implementation is shown in Fig.~\ref{atlas-7-lomass}. 
The data used the invariant mass region of 26 -- 66 GeV for the higher luminosity whereas less precise data extend this region 
down to 12 GeV. Here, one starts to observe a minor deviation between predictions of BS15 
and the other two PDFs. While the results of MHHT and CT models are closer to data, the BS15 predictions are still within the 
experimental error bars.
\begin{figure}[h!]
\begin{center}
\includegraphics[width=10.0cm]{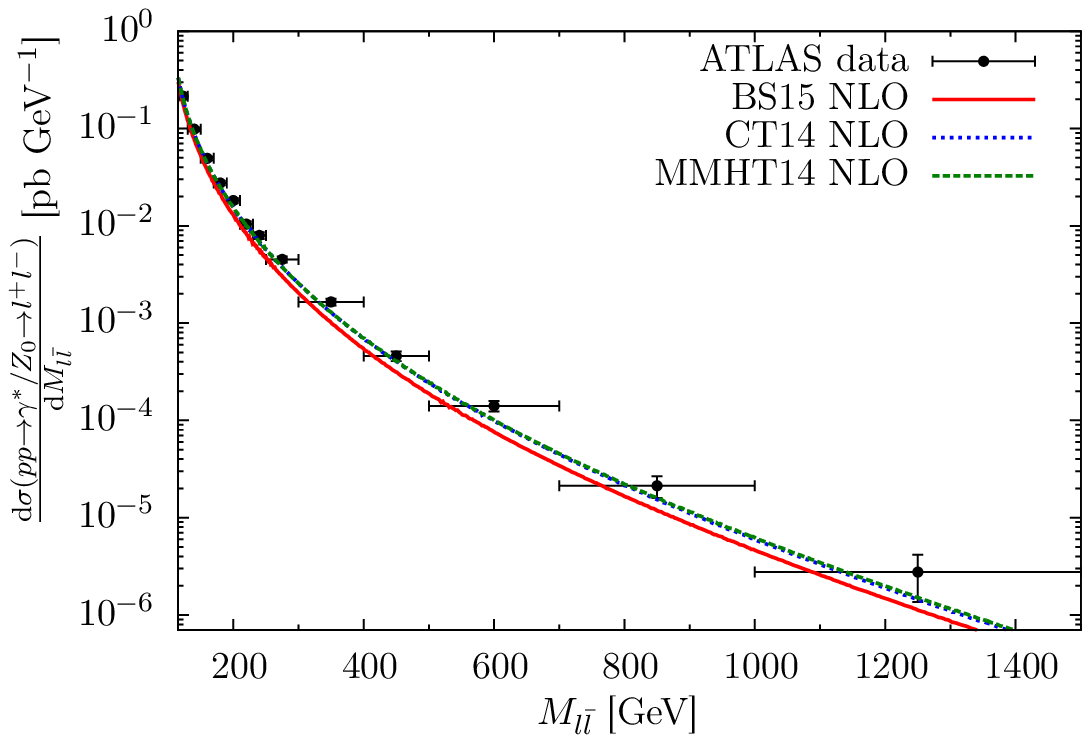}
\caption[*]{\baselineskip 1pt
The DY differential cross section measured in the 
di-electron channel by ATLAS at $\sqrt{s}$ = 7 TeV over the high-mass range 
116 -- 1500 GeV \cite{atlas13_highmass} vs QCD NLO predictions obtained by using 
PDF models from Refs.~\cite{mmht14,ct14,bs15}.}
\label{atlas-7-himass}
\end{center}
\end{figure}

We also selected the data for the high mass DY electron-positron pair production measured by the ATLAS Collaboration 
\cite{atlas13_highmass} which correspond to 4.9 fb$^{-1}$ luminocity within the mass range of 
$116 < M_{e^+e^-} < 1500$ GeV and in the kinematics defined by di-electron pairs with $p_\perp > 25$ GeV and $|\eta| < 2.5$. 
In Fig.~\ref{atlas-7-himass} the high mass distribution of $e^+e^-$ pairs is compared with the corresponding NLO pQCD predictions. 
The CT14 and MMHT14 results are pretty close to each other and show a very good agreement with data, while the BS15 result is somewhat 
below the data but is still within the error bars.
\begin{figure}[h!]
\begin{center}
\includegraphics[width=6.5cm]{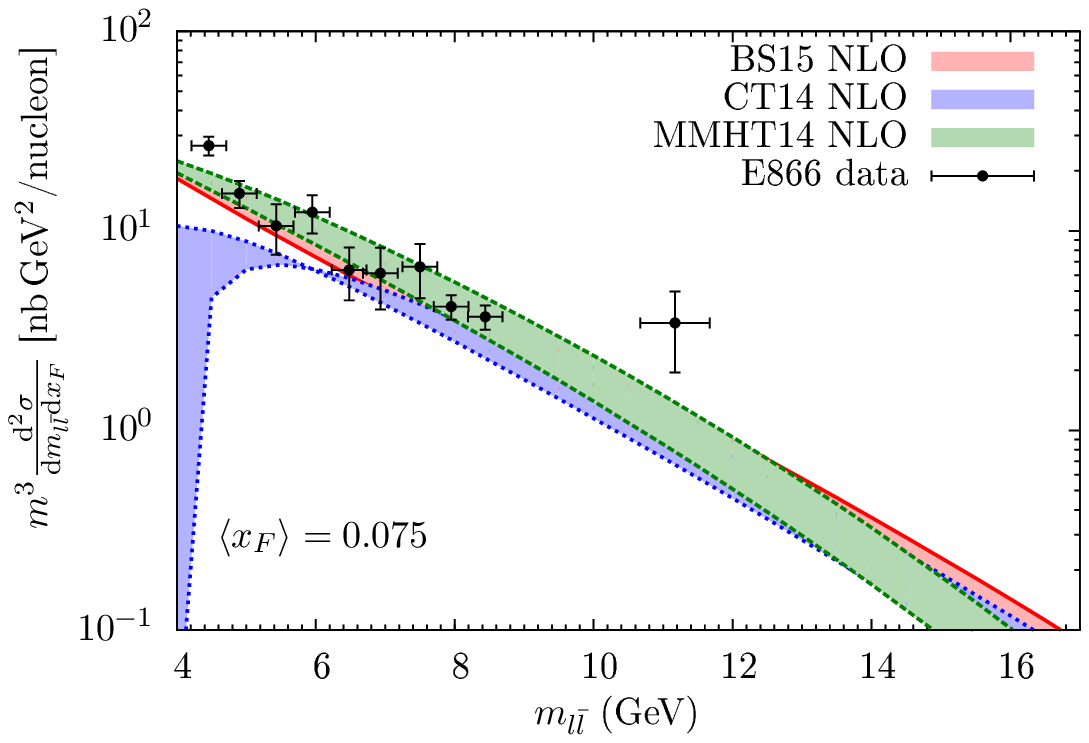}
\includegraphics[width=6.5cm]{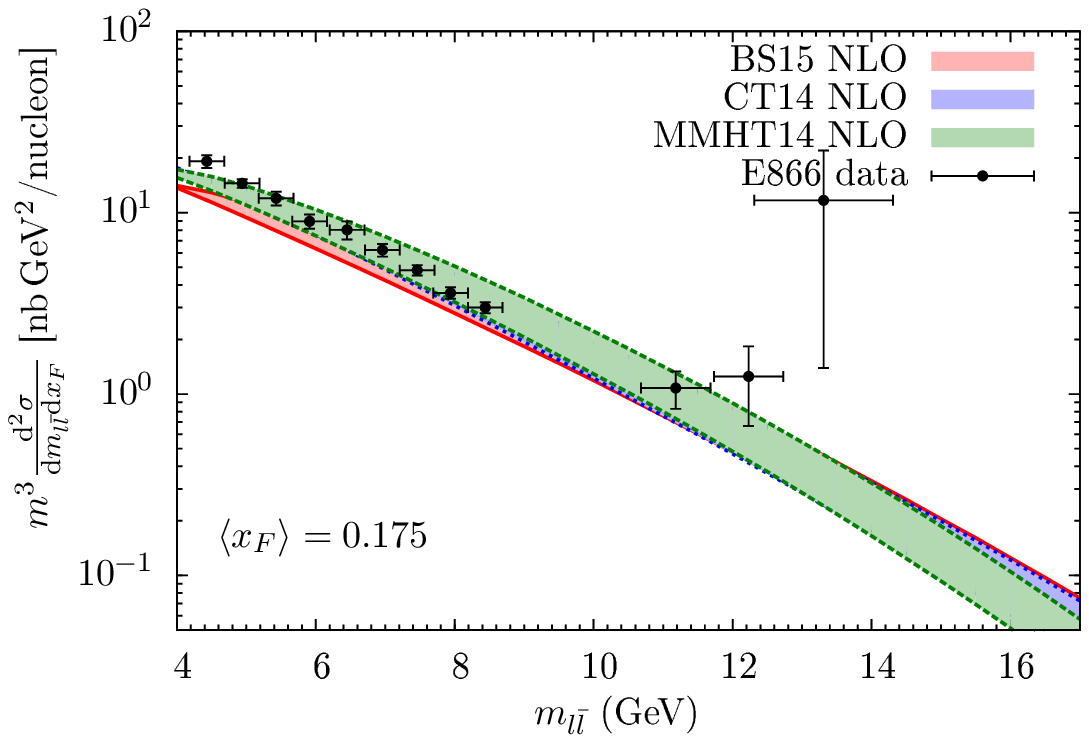}
\includegraphics[width=6.5cm]{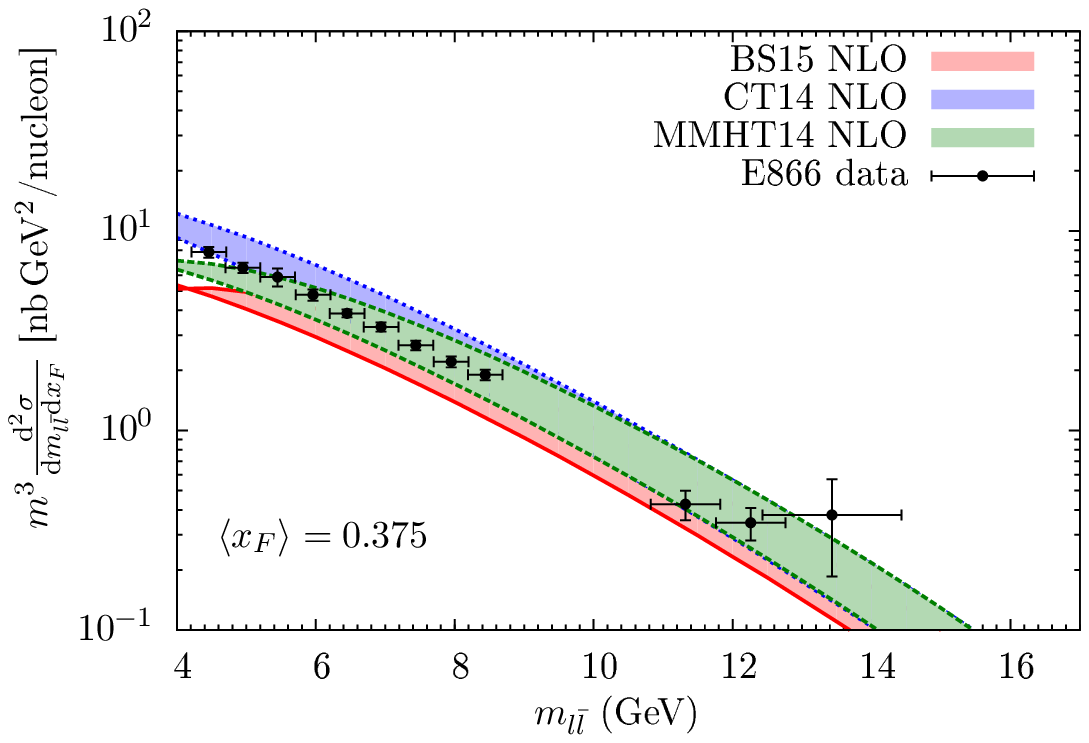}
\includegraphics[width=6.5cm]{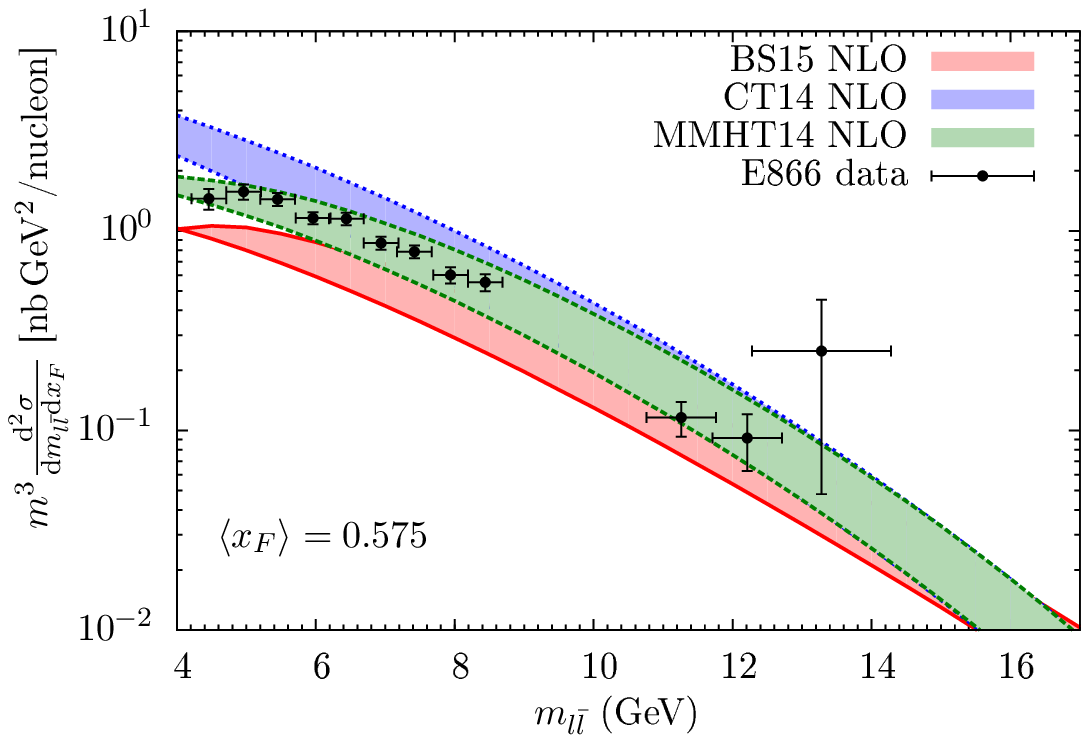}
\includegraphics[width=6.5cm]{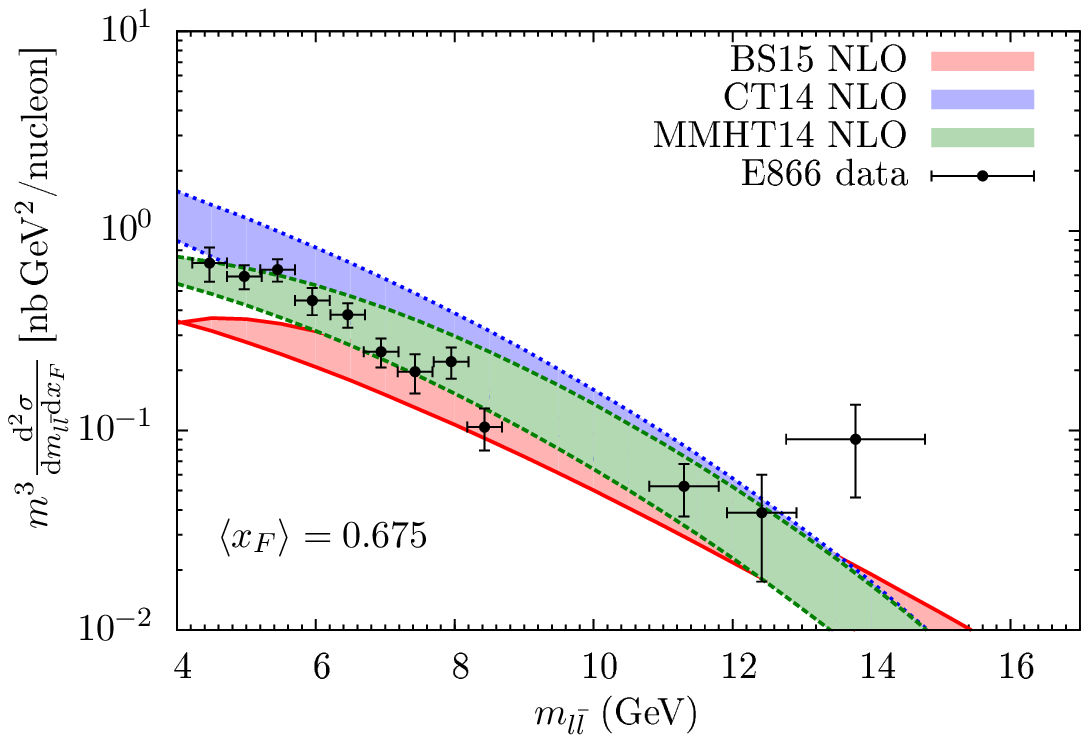}
\includegraphics[width=6.5cm]{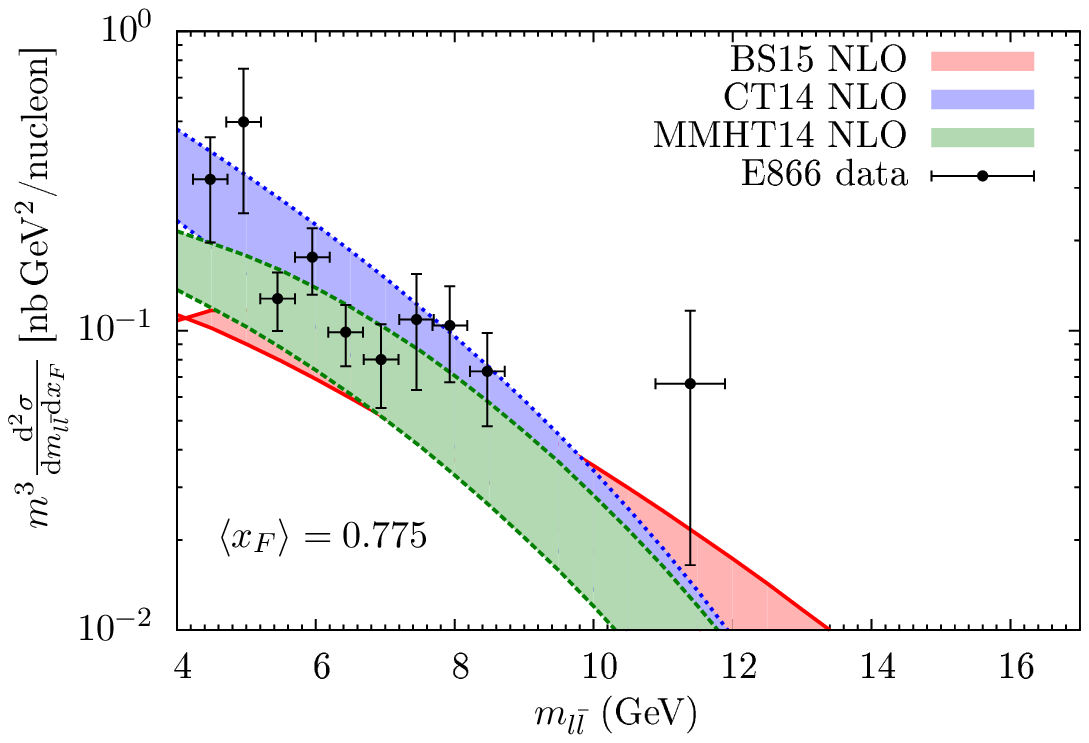}
\caption[*]{\baselineskip 1pt
The DY differential cross section measured in the di-muon channel by the E866 Collaboration
at $\sqrt{s}$ = 38.8 GeV \cite{e866} over a low-mass range and various Feynman $\langle x_F \rangle$ variables 
vs QCD NLO predictions obtained by using PDF models from Refs.~\cite{mmht14,ct14,bs15}. The uncertainty
w.r.t. scale variations within the interval $M_{l\bar l}/2<\mu<2M_{l\bar l}$ is shown as filled error bands.}
\label{e866}
\end{center}
\end{figure}

At much lower energies at FNAL $\sqrt{s}$ = 38.8 GeV the E866/NuSea Collaboration provided the DY differential 
distributions over a broad kinematics range \cite{e866}. In this case, we have implemented the formalism by 
Sutton et al. \cite{K-factor} where both NLO corrections for Compton and annihilation subprocesses were included.
This formalism results in a fair description of the E866/NuSea data on DY differential distributions in the proton-proton 
collisions for three PDF models (BS15, CT14 and MMHT14) at various $\langle x_F \rangle$ and dilepton masses 
shown in Fig.~\ref{e866}. Here, an overall uncertainty w.r.t. scale variations within the interval 
$M_{l\bar l}/2<\mu<2M_{l\bar l}$ is shown as filled error bands. Note, the uncertainties grow with $\langle x_F \rangle$ 
similarly in all three models. Noticeable deviations between the predictions from distinct PDF sets emerge mostly 
at small invariant masses $M_{l\bar l}<6$ GeV where some deficiency of the predictions can be observed. 
Specifically for this case, one observes that the CT14 PDF is rather unstable at $\langle x_F \rangle = 0.075$ 
and the CT14/BS15 PDFs do not work well at $\langle x_F \rangle \simeq 0.5-0.7$, while MMHT14 demonstrates
a better description in both normalisation and shape. For intermediate values of the Feynman variable and large 
$M_{l\bar l}$ all the considered PDFs provide a fairly good description of the low energy data.

\section{$Z/\gamma^*$ boson rapidity distribution}

The events and analyses for this and further sections were generated with Pythia v8 \cite{pythia8} 
and Rivet \cite{rivet2}. For the $Z/\gamma^*$ boson rapidity distribution the standard LO Pythia 
(with parton showers and NLO PDFs) was employed. The parton shower within Pythia effectively 
resums the higher-order real corrections due to emissions off the colored initial states. The latter 
corrections significantly affect the kinematics of the final states and is need for a proper 
treatment of the differential distributions. Therefore, this approach differs from the full NLO 
calculation by the virtual corrections only. However, these corrections can only affect the overall 
normalisation of the cross sections and are typically accounted for by a universal $K$-factor 
which is practically independent on kinematics of the final states. Since the experimental data 
for the Drell-Yan process at Tevatron and LHC are typically provided in terms of ratios of the 
differential-to-total cross sections, one does not need to explicitly compute the virtual corrections 
since they cancel out in these ratios and do not affect our conclusions. Besides, other model 
uncertainties such as the scale uncertainty mostly cancel in such ratios as well. So it is not 
surprising that this method leads to a very good agreement of various PDF models with 
the Tevatron and LHC data on the ratios in the whole kinematic region (see below).
\begin{figure}[h!]
\begin{center}
\includegraphics[width=8.0cm]{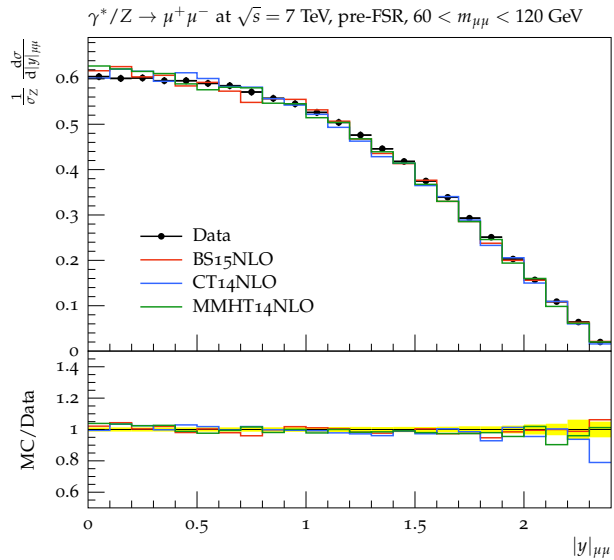}
\caption[*]{\baselineskip 1pt
The DY differential cross section measured in the di-muon channel by CMS at
$\sqrt{s}$ = 7 TeV \cite{cms13} vs QCD NLO predictions obtained by using 
PDF models from Refs.~\cite{mmht14,ct14,bs15}.}
\label{cms15}
\end{center}
\end{figure}

\begin{figure}[h!]
\begin{center}
\includegraphics[width=8.0cm]{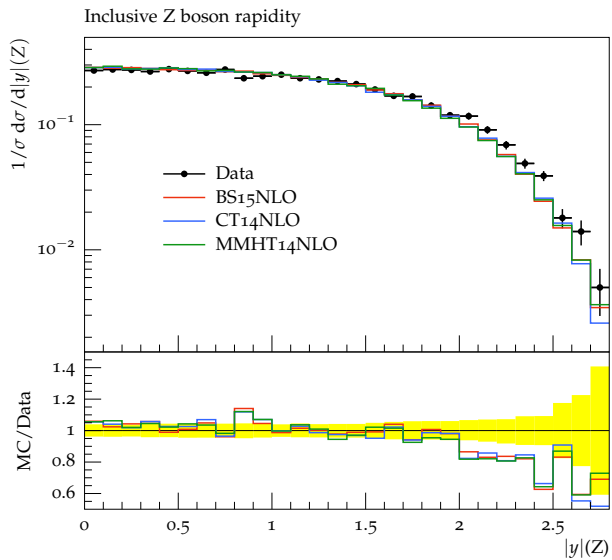}
\caption[*]{\baselineskip 1pt
The DY differential cross section measured in the di-electron channel by the D0 Collaboration
at $\sqrt{s}$ = 1.96 TeV \cite{d0-07} vs QCD NLO predictions obtained by using 
PDF models from Refs.~\cite{mmht14,ct14,bs15}.}
\label{d07}
\end{center}
\end{figure}

For validation of the considered PDF models and theoretical methods, we display in Fig. \ref{cms15} the results on the 
rapidity distribution from the CMS Collaboration at $\sqrt{s}$ = 7 TeV \cite{cms13} 
corresponding to the integrated luminosity of 4.5(4.8) fb$^{-1}$. The di-muon channel was used for rapidity 
reconstruction within certain invariant mass intervals, from which we have selected the one corresponding 
to a vicinity of the $Z$ peak, $60<m_{\mu\bar{\mu}}<120$ GeV.
Analogically, the rapidity distributions from the D0 Collaboration
at a lower energy $\sqrt{s}$ = 1.96 TeV \cite{d0-07} in the di-electron channel 
and the mass range 71 $\leq \mbox{M}_{e^+e^-} \leq $ 111 GeV corresponding to 
an integrated luminosity of 0.4fb$^{-1}$ were compared to the theoretical predictions 
with BS, MHHT and CT models. The corresponding results are shown in Fig.~\ref{d07}.
All the NLO pQCD predictions exhibit a fair agreement with the existing data from both LHC 
and Tevatron measurements.

\section{$Z/\gamma^*$ boson transverse momentum spectrum}

\begin{figure}[h!]
\begin{center}
\includegraphics[width=8.0cm]{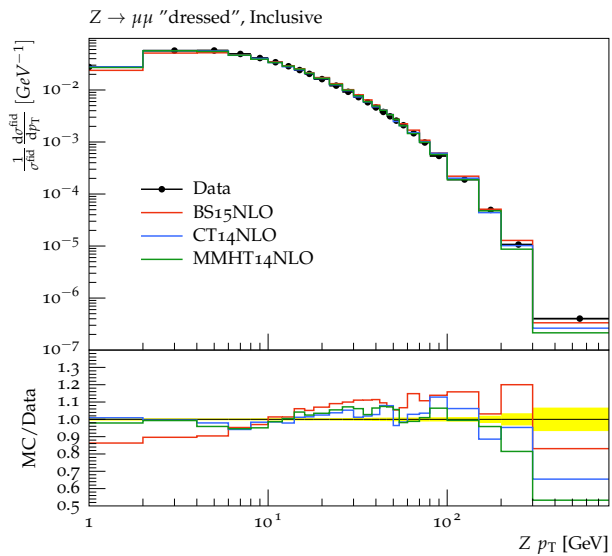}
\caption[*]{\baselineskip 1pt
The ATLAS 7 TeV data on $Z$-boson transverse momentum distribution with 
the full detector acceptance \cite{atlas14} vs QCD NLO predictions obtained by
using PDF models from Refs.~\cite{mmht14,ct14,bs15}.}
\label{ZpT_ATLAS}
\end{center}
\end{figure}

\begin{figure}[hbt]
\begin{center}
\includegraphics[width=6.0cm]{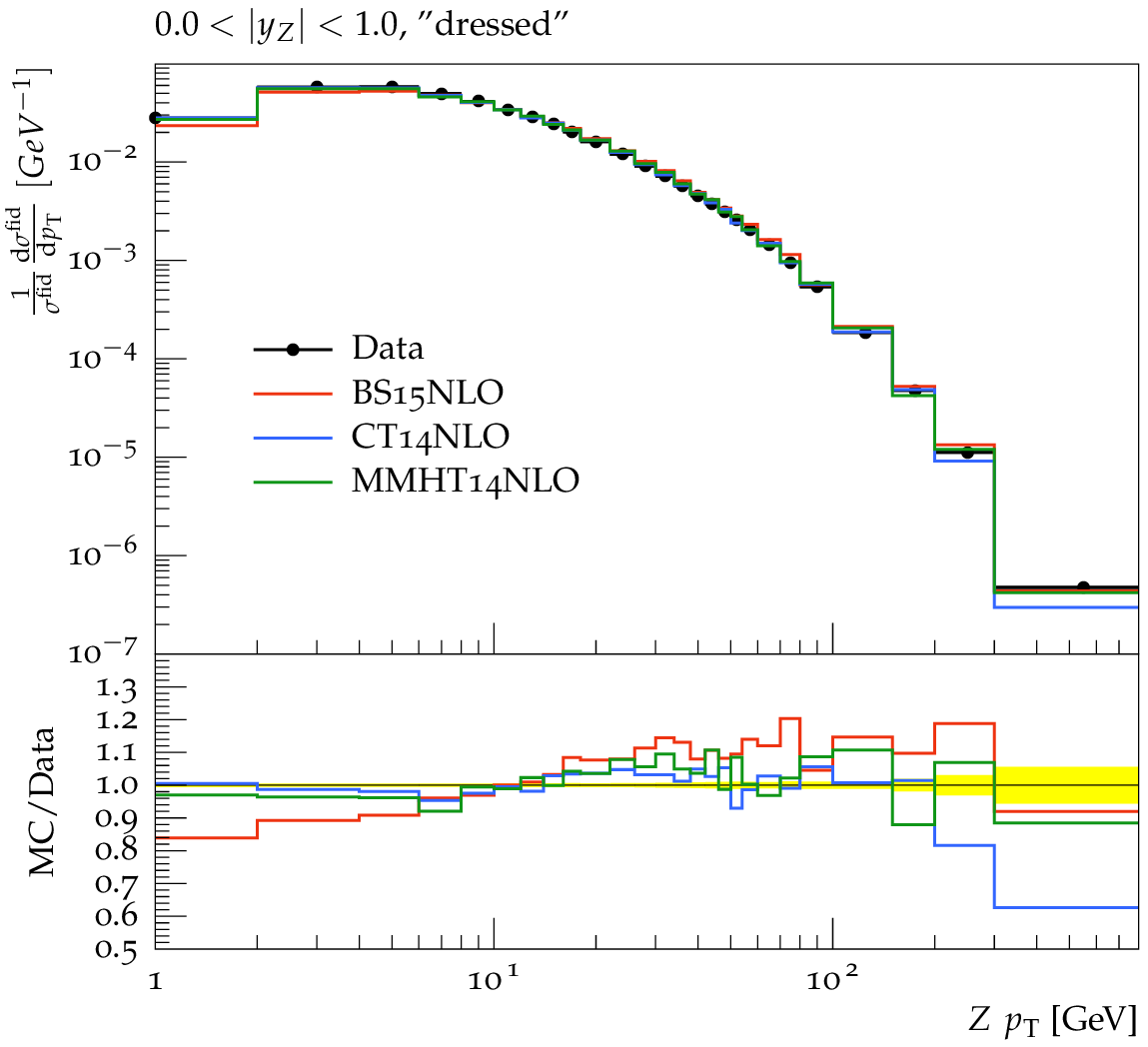}
\includegraphics[width=6.0cm]{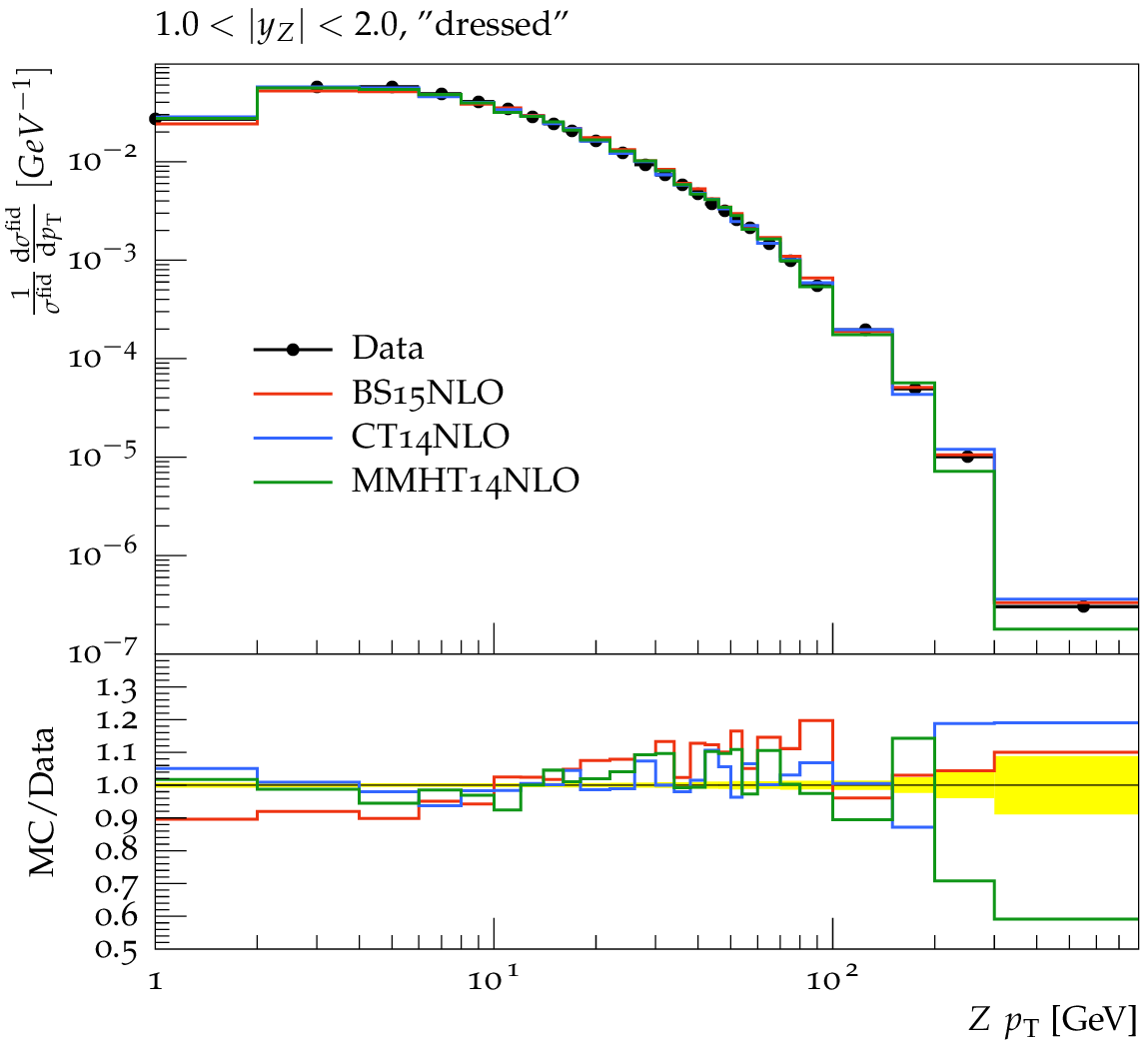}
\includegraphics[width=6.0cm]{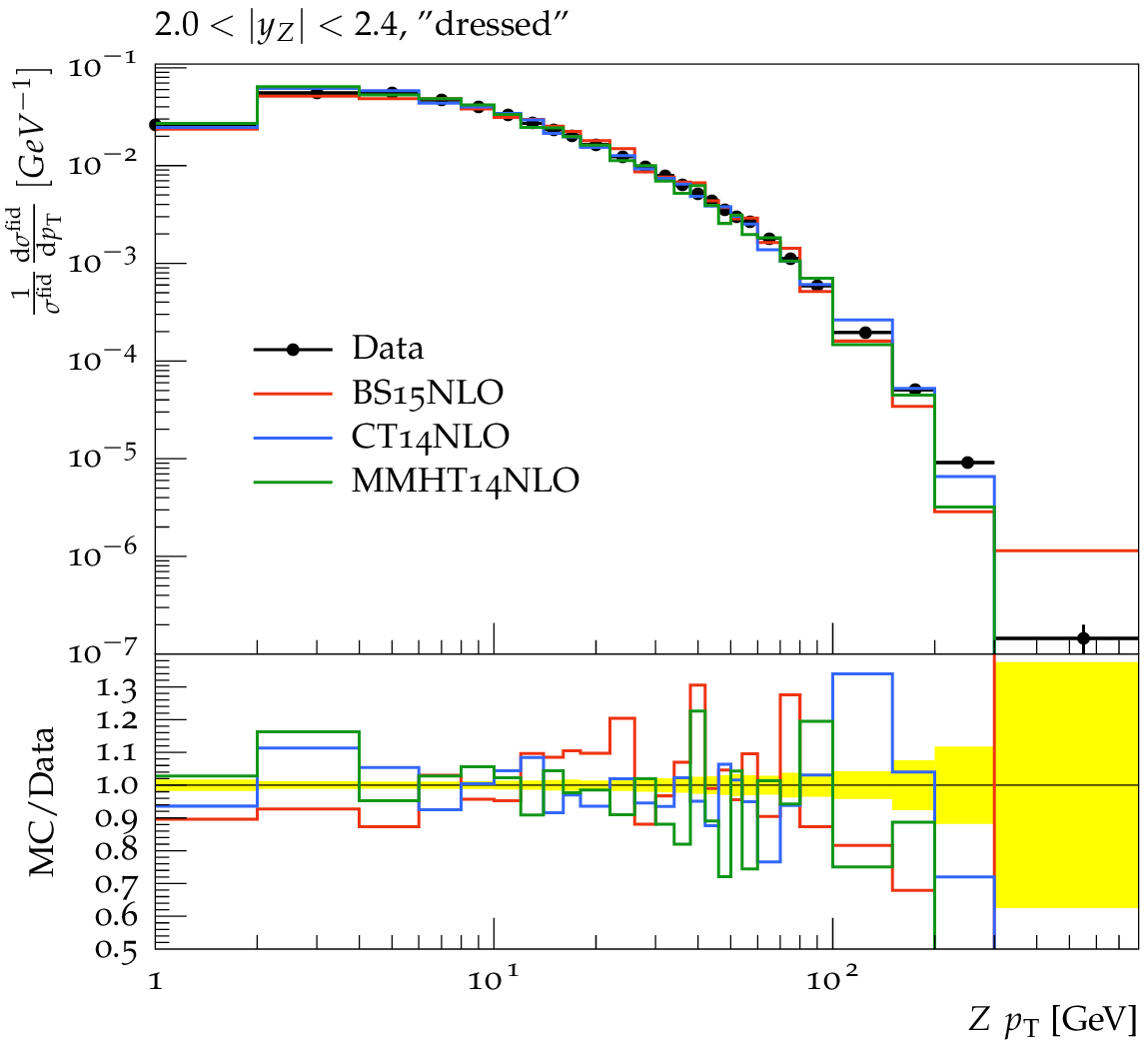}
\caption[*]{\baselineskip 1pt
The ATLAS 7 TeV data on $Z$-boson transverse momentum distribution 
for three different rapidity bins \cite{atlas14} vs QCD NLO predictions 
obtained by using PDF models from Refs.~\cite{mmht14,ct14,bs15}.}
\label{ZpT_ATLAS_forward}
\end{center}
\end{figure}

A high-precision data on the $Z/\gamma^*$ boson transverse momentum distribution
up to 800 GeV has been presented by ATLAS in Ref.~\cite{atlas14} at $\sqrt{s}$ = 7 TeV corresponding to 
the integrated luminosity of 4.7 fb$^{-1}$. In Fig.~\ref{ZpT_ATLAS} we consider the full detector acceptance 
whereas the $Z$-boson transverse momentum for three different 
rapidity bins is shown in Fig.~\ref{ZpT_ATLAS_forward}. A good description was obtained as well including 
rather high-$p_\perp$ region where in principle the higher order corrections could start to play an important role, 
as well as the statistical fluctuations for the forward region ($2<|y_Z|<2.4$) could become noticeable 
due to a smaller number of events generated in the forward/high-$p_T$ kinematics. A disagreement 
of the model predictions in the highest $p_\perp$ bins, besides poor statistics, could be due to the fact we employ the simplistic 
analysis with LO matrix elements and NLO PDFs (plus parton shower) and not the full-precision higher-order $Z + jets$ matrix elements. 
Remarkably enough, one can see that the PDFs tested here and our simplistic LO analysis give a fair description of data for almost entire range 
of $Z$-boson $p_T$'s.

\begin{figure}[hbt]
\begin{center}
\includegraphics[width=6.0cm]{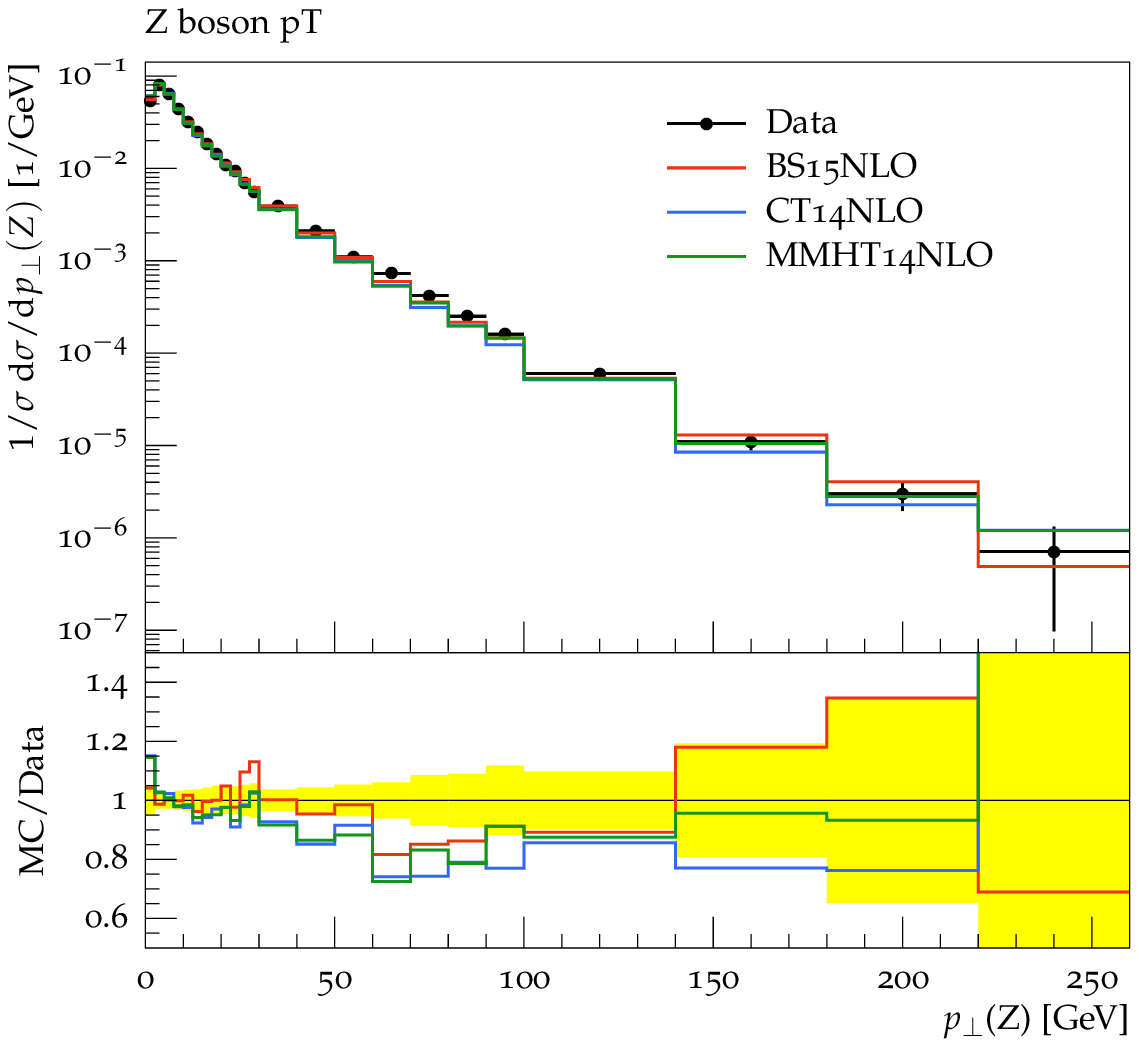}
\includegraphics[width=6.0cm]{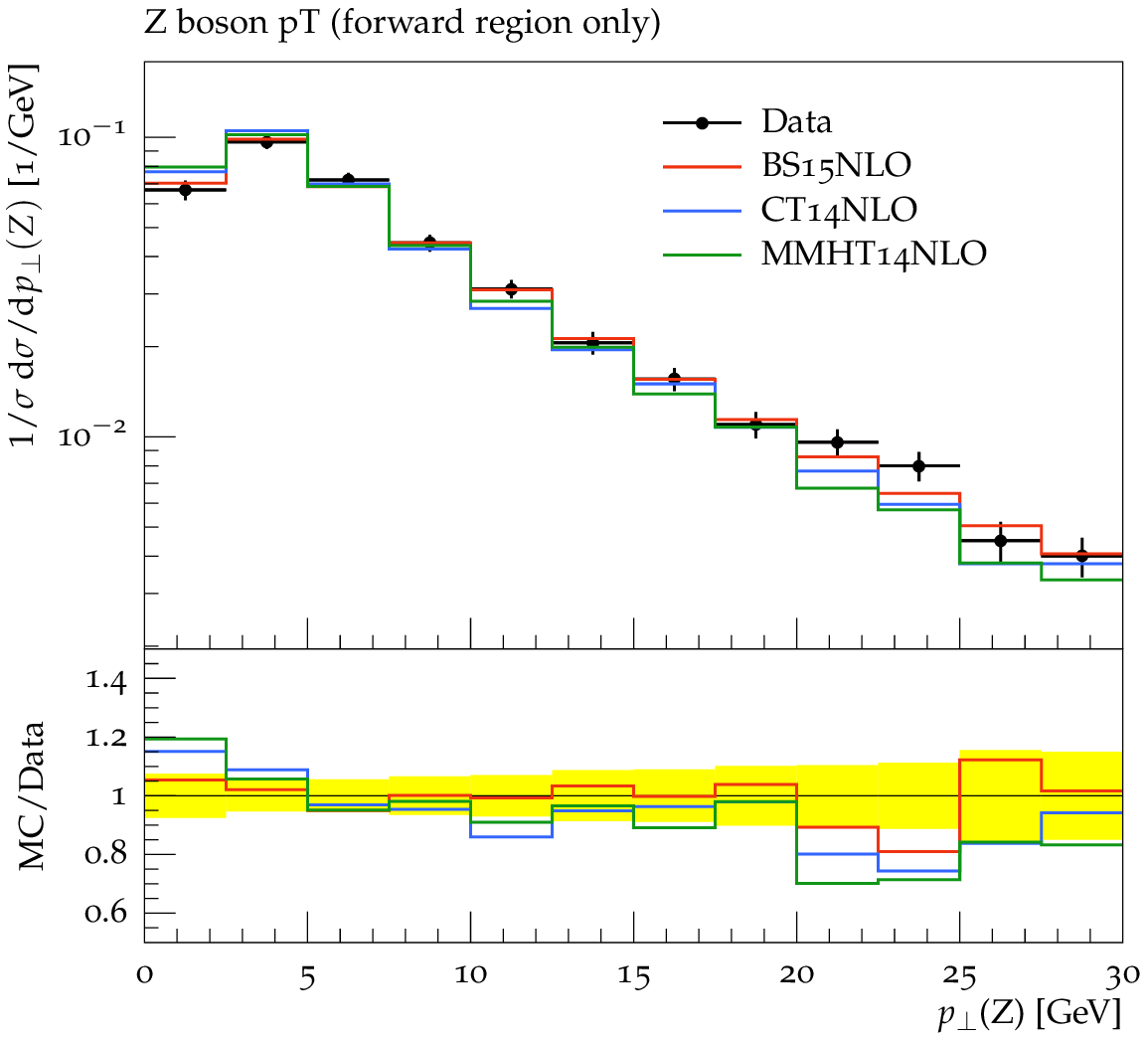}
\caption[*]{\baselineskip 1pt
Data from D0 \cite{d0-08} at 1.98 TeV on $Z$-boson transverse momentum distribution for 
two kinematic regions vs QCD NLO predictions obtained by using PDF models from Refs.~\cite{mmht14,ct14,bs15}.}
\label{ZpT_D0}
\end{center}
\end{figure}

It is worth to present similar results for the $Z/\gamma^*$ boson transverse momentum distribution 
from the D0 Collaboration at Tevatron \cite{d0-08} in $p\bar p$ collisions at $\sqrt{s}$ =1.96 TeV in the di-electron channel 
corresponding to the integrated luminosity of 0.98 fb$^{-1}$. In Fig.~\ref{ZpT_D0} the $Z$-boson transverse momentum 
distribution is plotted in two regions, the one that covers the full detector acceptance and the one which is focused on 
forward production, \textit{i.e.}, for $\eta > 2$ and $p_\perp<30$ GeV. Similarly to the LHC case, the theoretical modelling
with the chosen PDFs exhibit rather good description of the Tevatron data within the experimental uncertainties.

\section{Concluding remarks}

To summarize, the NLO pQCD predictions with the MHHT14, CT14 and BS15 PDFs studied in this work have resulted in 
a fair description of a broad range of DY data from FNAL-NUSEA energies up to LHC energies while a few minor deviations, 
mainly at the edges of the respective phase space, have been observed. Given the fact that three very different PDF sets 
lead to rather similar results we conclude that the NLO BS15 model having a much fewer free parameters which are fitted 
to the DIS data only is not worse than the other most recent parameterisations and should be considered on the same footing 
as the current global PDF fits when it comes to the DY process at high energies.
\begin{figure}[!h]  
\begin{center}
\includegraphics[width=4.45cm]{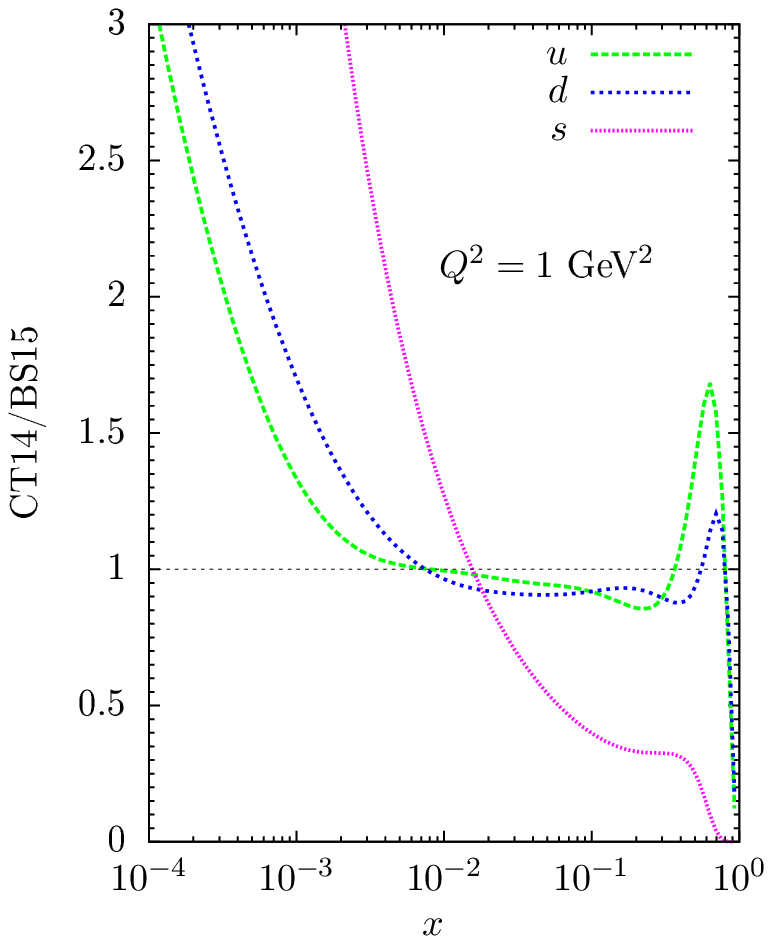}
\includegraphics[width=4.45cm]{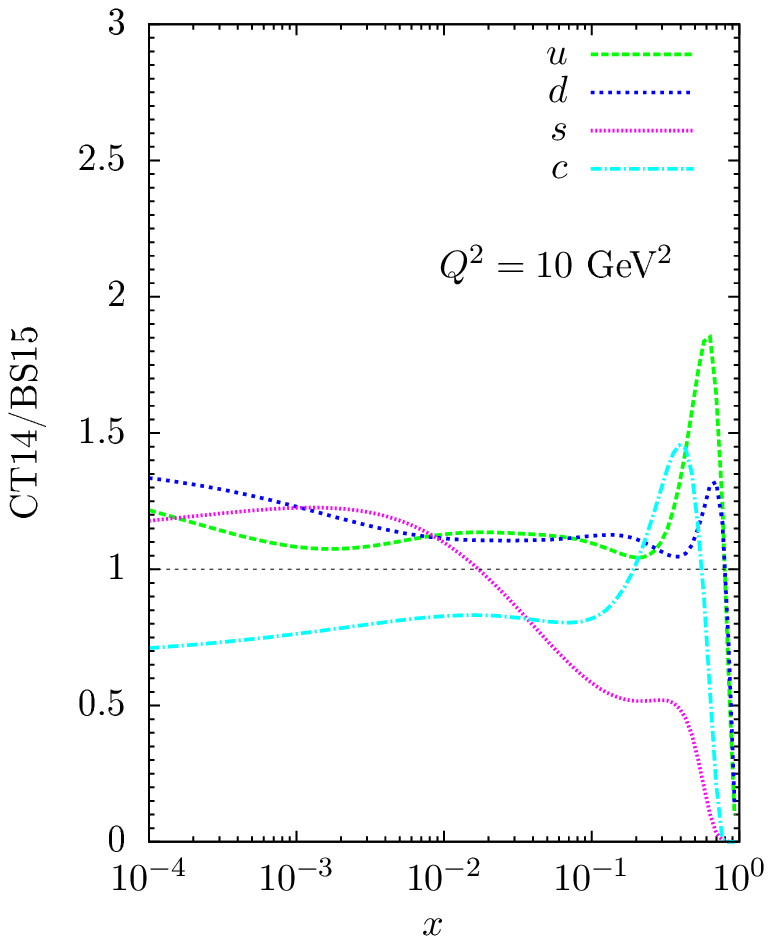}
\includegraphics[width=4.45cm]{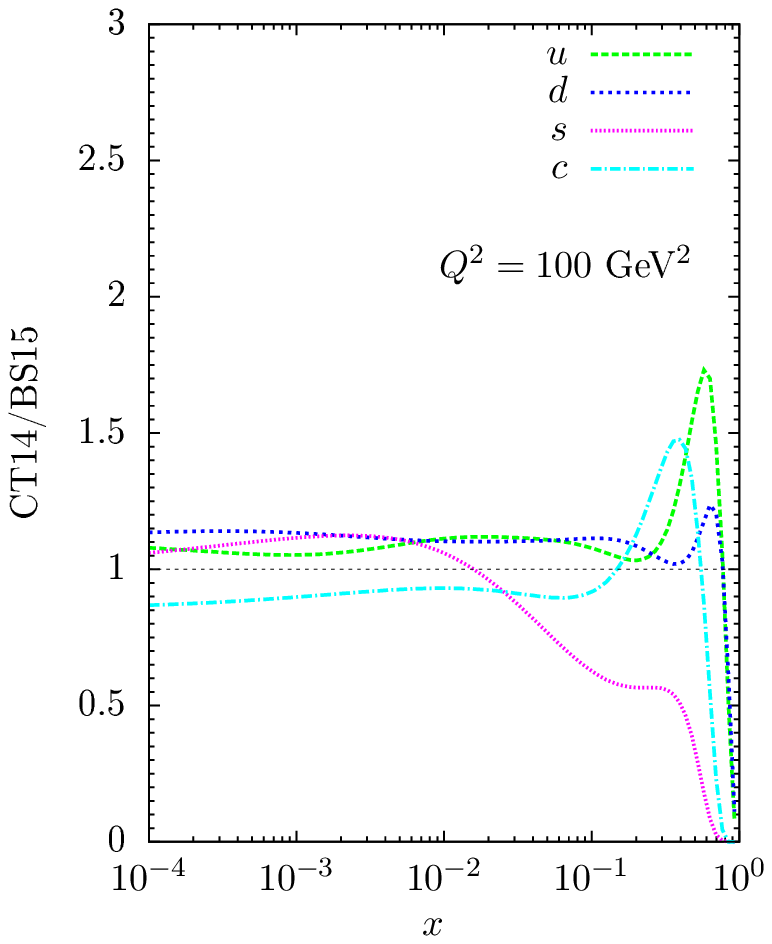}
\includegraphics[width=4.45cm]{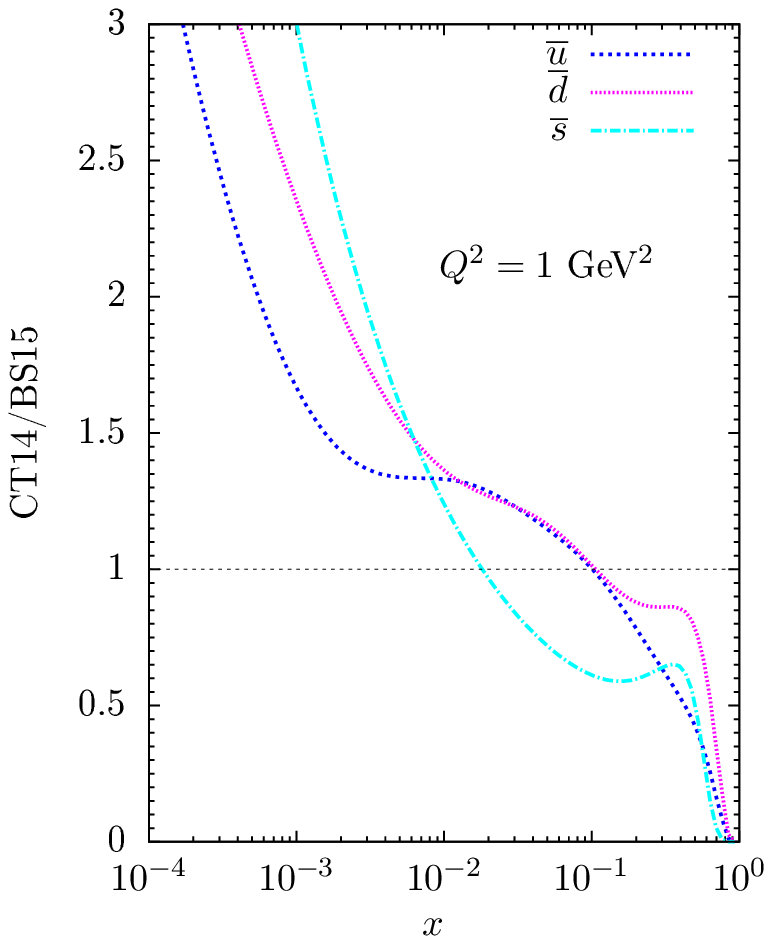}
\includegraphics[width=4.45cm]{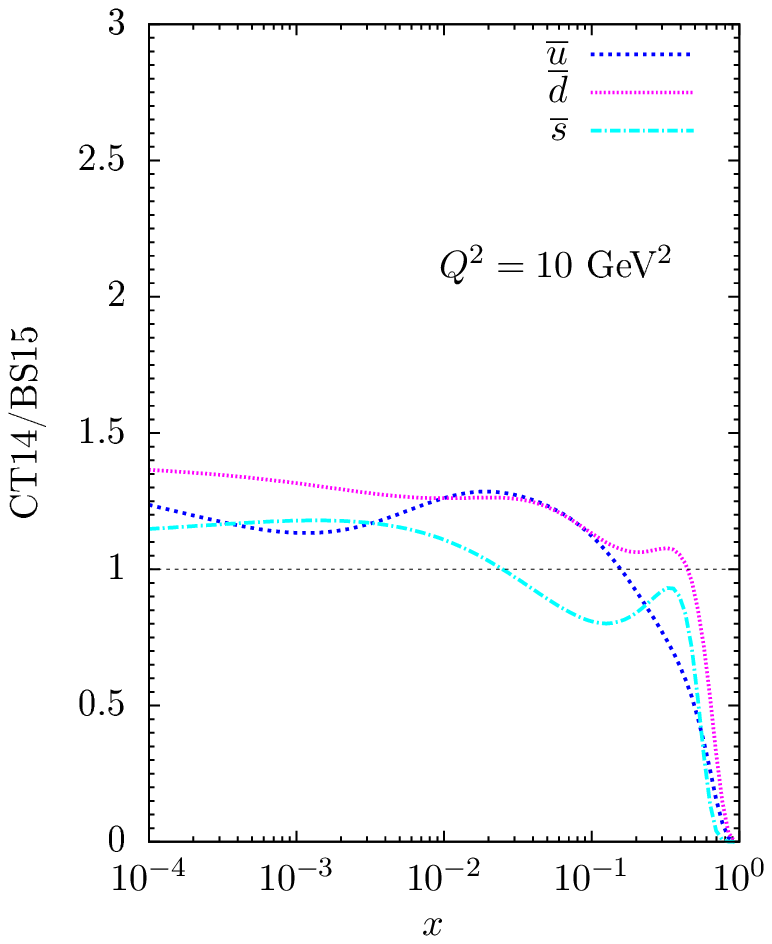}
\includegraphics[width=4.45cm]{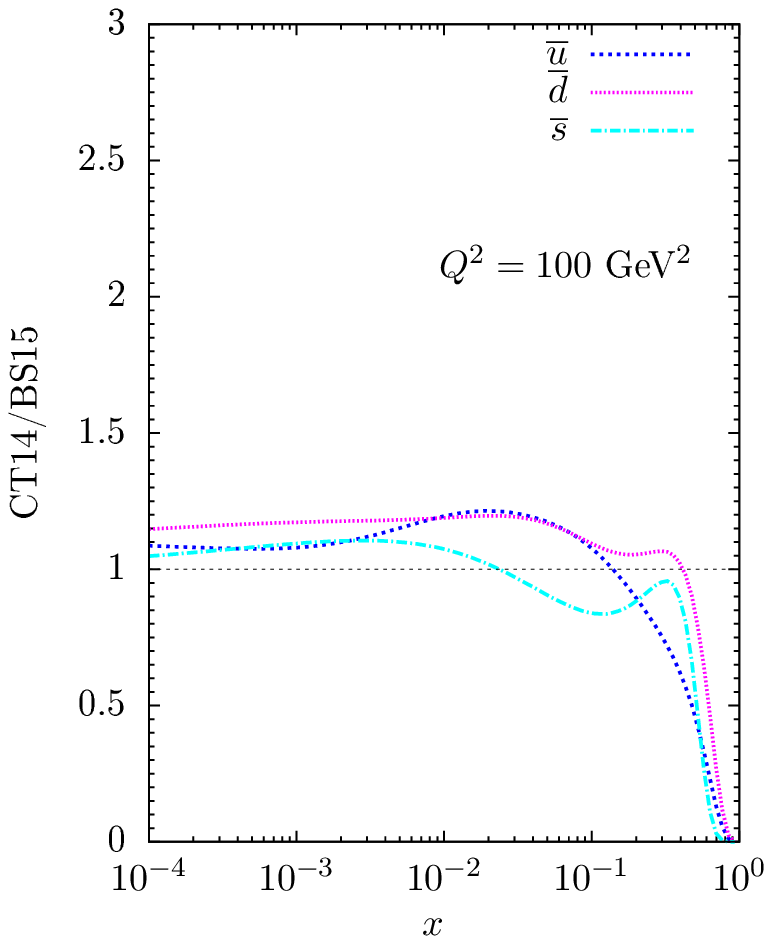}
\caption[*]{\baselineskip 1pt 
The CT14-to-BS15 ratios for quark (upper row) and antiquark (lower row) NLO PDFs 
as functions of $x$ and $Q^2=1,\,10,\,100$ GeV$^2$.}
\label{ratios}
\end{center}
\end{figure}

\begin{figure}[!h]
\begin{center}
\includegraphics[width=6.0cm]{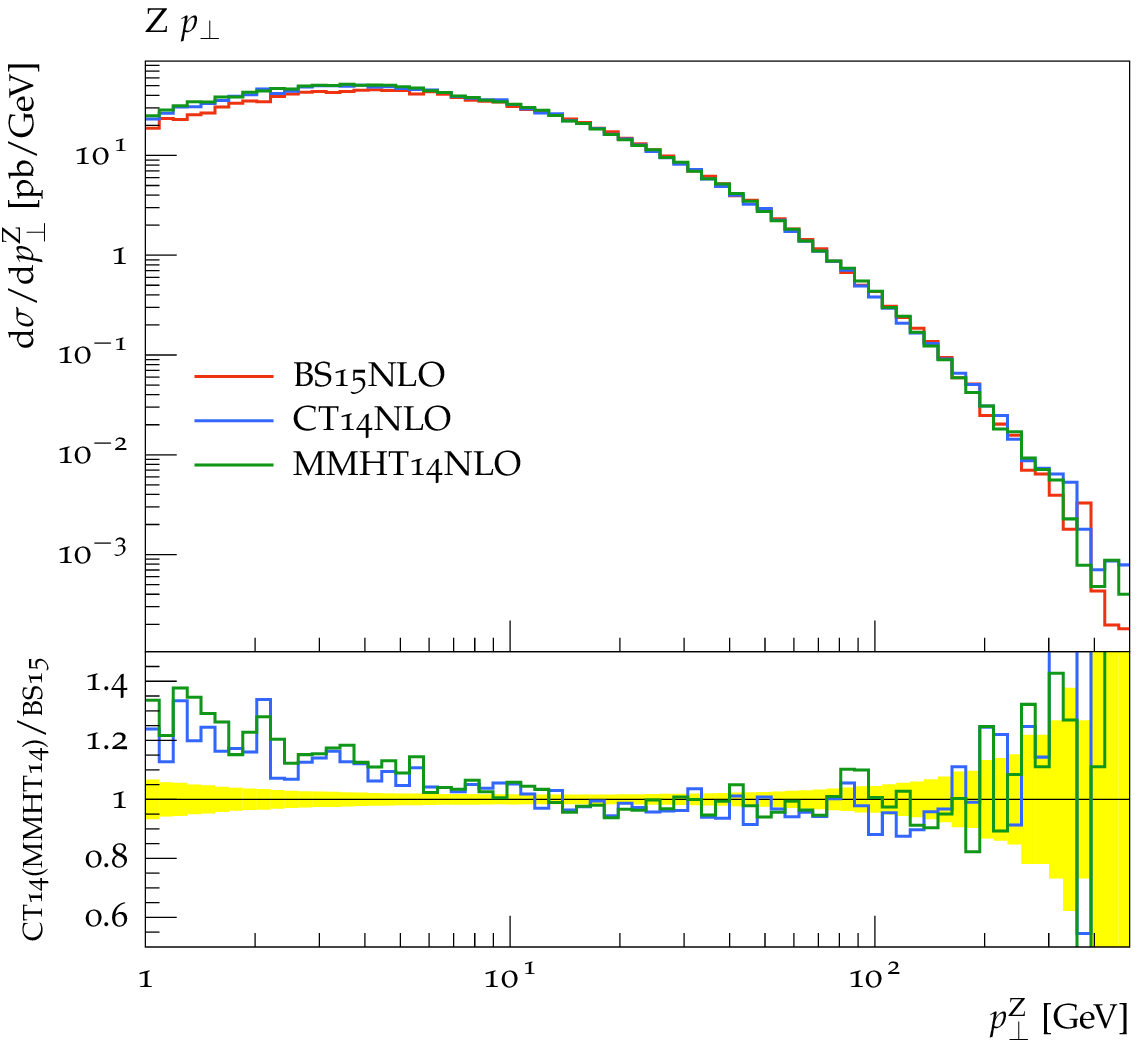}
\includegraphics[width=6.0cm]{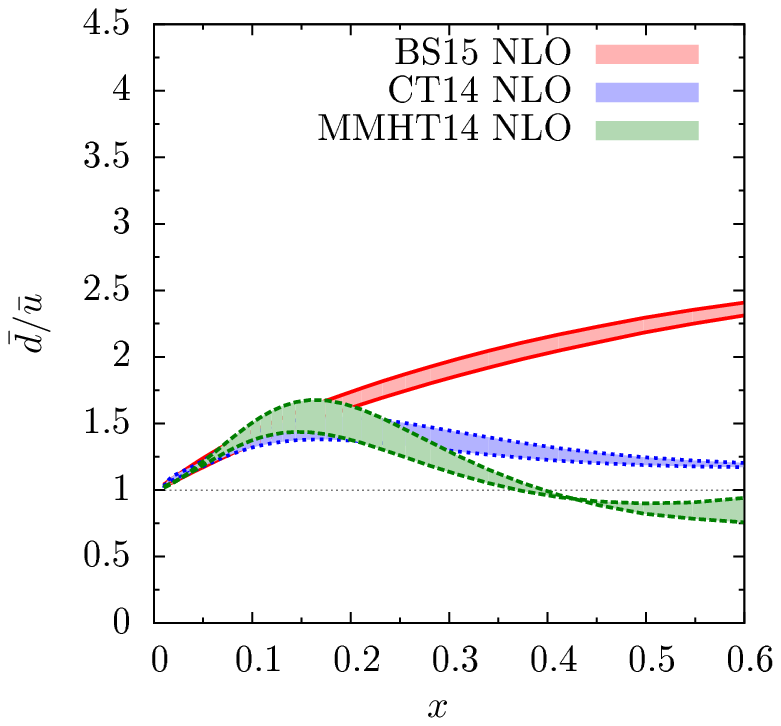}
\caption[*]{\baselineskip 1pt
Predictions for the $Z$ boson transverse momentum distribution 
at $\sqrt{s}=13$ TeV (left panel) and for the ratio of sea quark PDFs 
$\bar{d}/\bar{u}$ as a function of fraction $x$ with scale variations within 
$1 < Q^2 < 100$ GeV$^2$ interval (right panel) obtained by using PDF 
models from Refs.~\cite{mmht14,ct14,bs15}.}
\label{ZINC}
\end{center}
\end{figure}

Extrapolating to the Drell-Yan process at even larger energies of LHC Run II ($\sqrt{s}=13$ TeV) we do not expect more significant 
differences between the BS15 predictions and those of other PDF models since typical $x$ values we probe 
there get even smaller at higher energies. In Fig.~\ref{ratios} we present (anti)quark PDF CT14-to-BS15 ratios 
(central values) as functions of $x$ and $Q^2=1,\,10,\,100$ GeV$^2$. This figure demonstrates that 
the differences between the considered PDFs becomes smaller at larger $Q^2$ and $x\lesssim 10^{-2}$ typical 
for Drell-Yan at large $M_{\bar ll}$ at the LHC; inclusion of error bars does not change this picture. 
Similar situation holds for comparison of BS15 with the MHHT14 PDFs. As an illustration, in Fig.~\ref{ZINC} (left)
we show the transverse momentum distributions of the $Z$-boson at $\sqrt{s}=13$ TeV 
for BS15, CT14 and MHHT14 PDFs. A noticeable deviation of the BS15 prediction, which lies below the other PDFs, is seen 
only at low $p_\perp^Z<10$ GeV, otherwise differences between the predictions for this and other observables 
such as $y$ and $M_{\bar ll}$ distributions are small. We conclude that the BS15 NLO PDF is a good tool 
to investigate pQCD physics in the second run of LHC measurements. 

As is seen in Fig.~\ref{e866}, at low energies the PDF models exhibit more substantial differences in shapes of the invariant 
mass and $x_F$ distributions than those at high energies. This fact indicates the critical importance of low energy Drell-Yan 
measurements such as the FNAL E906/SeaQuest experiment \cite{Reimer:2011zza}. In particular, the E906 experiment 
will determine the ratio $\bar{d}/\bar{u}$ at large $x$ (up to $x\sim 0.5-0.6$). The statistical BS15 model predicts 
$\bar{d}/\bar{u}>1$ at large $x$ as illustrated in Fig.~\ref{ZINC} (right) together with CT14 and MMHT14 predictions where 
the filled bands correspond to scale variations within $1 < Q^2 < 100$ GeV$^2$ interval for each PDF model.
This characteristic effect is based on the Pauli exclusion principle and the fact that in the proton there are two $u$-quarks 
and one $d$-quark such that more $\bar{d}$ compared to $\bar{u}$. A high precision measurement of such a ratio would 
therefore become a crucial test of physics behind the PDF models.

\vskip0.2cm
{\bf Acknowledgments}

Useful discussions with Gunnar Ingelman, Jesper Roy Christiansen and Torbjorn Sj\"ostrand are gratefully acknowledged. 
E. B. is funded by CAPES and CNPq (Brazil), contract numbers 2362/13-9 and 150674/2015-5.
R. P. is supported by the Swedish Research Council, contract number 621-2013-428.

\end{document}